\documentclass[pdflatex,sn-mathphys-num]{sn-jnl}

\usepackage{graphicx}%
\usepackage{multirow}%
\usepackage{amsmath,amssymb,amsfonts}%
\usepackage{amsthm}%
\usepackage{mathrsfs}%
\usepackage[title]{appendix}%
\usepackage{xcolor}%
\usepackage{textcomp}%
\usepackage{manyfoot}%
\usepackage{booktabs}%
\usepackage{algorithm}%
\usepackage{algorithmicx}%
\usepackage{algpseudocode}%
\usepackage{listings}%
\usepackage[separate-uncertainty = true]{siunitx}

\begin{document}

\title[Cryogenic Feedforward of a Photonic Quantum State]{Cryogenic Feedforward of a Photonic Quantum State}


\author*[1,2]{\fnm{Frederik} \sur{Thiele}}\email{frederik.thiele@uni-paderborn.de}
\equalcont{These authors contributed equally to this work.}

\author[2]{\fnm{Niklas} \sur{Lamberty}}
\equalcont{These authors contributed equally to this work.}

\author[1]{\fnm{Thomas} \sur{Hummel}}

\author[1,2]{\fnm{Nina A.} \sur{Lange}}

\author[1,2]{\fnm{Lorenzo M.} \sur{Procopio}}

\author[1,2]{\fnm{Aishi} \sur{Barua}}

\author[3]{\fnm{Sebastian} \sur{Lengeling}}

\author[1]{\fnm{Viktor} \sur{Quiring}}

\author[1]{\fnm{Christof} \sur{Eigner}}

\author[3]{\fnm{Christine} \sur{Silberhorn}}

\author[1,2]{\fnm{Tim J.} \sur{Bartley}}

\affil[1]{\orgdiv{Institute for Photonic Quantum Systems (PhoQS)}, \orgname{Paderborn University}, \orgaddress{\street{Warburger Str. 100}, \city{Paderborn}, \postcode{33098},  \country{Germany}}}

\affil[2]{\orgdiv{Department of Physics}, \orgname{Paderborn University}, \orgaddress{\street{Warburger Str. 100}, \city{Paderborn}, \postcode{33098},  \country{Germany}}}

\affil[3]{\orgdiv{Integrated Quantum Optics Group, Institute for Photonic Quantum Systems (PhoQS)}, \orgname{Paderborn University}, \orgaddress{\street{Warburger Str. 100}, \city{Paderborn}, \postcode{33098},  \country{Germany}}}


\abstract{Modulation conditioned on measurements on entangled photonic quantum states is a cornerstone technology of optical quantum information processing. Performing this task with low latency requires combining single-photon-level detectors with both electronic logic processing and optical modulation in close proximity. In the technologically relevant telecom wavelength band, detection of photonic quantum states is best performed with high-efficiency, low-noise, and high-speed detectors based on the photon-induced breakdown of superconductivity. Therefore, using these devices for feedforward requires mutual compatibility of all components under cryogenic conditions. Here, we demonstrate low-latency feedforward using a quasi-photon-number-resolved measurement on a quantum light source. Specifically, we use a multipixel superconducting nanowire single-photon detector, amplifier, logic, and an integrated electro-optic modulator \emph{in situ} below \SI{4}{K}. We modulate the signal mode of a spontaneous parametric down-conversion source, conditional on a photon-number measurement of the idler mode, with a total latency of \SI{23(3)}{ns}. The photon-number discrimination actively manipulates the signal mode photon statistics, which is itself a central component in photonic quantum computing reliant on heralded single-photon sources. This represents an important benchmark for the fastest quantum photonic feedforward experiments comprising measurement, amplification, logic and modulation. This has direct applications in quantum computing, communication, and simulation protocols.}


\maketitle

\section{Introduction}\label{sec:intro}
Quantum photonics leverages measurements on quantum optical states to perform information processing tasks. In some tasks, such as generating shared secret keys~\cite{BB84}, or sampling from complex distributions~~\cite{Aaronson2011,Hamilton2017,Zhong2020,Madsen2022,Deng2023}, measuring the final state is sufficient. However, in other cases, such as linear optical quantum computing~\cite{Knill2001}, measurement-based quantum computing~\cite{Raussendorf2001} and quantum teleportation~\cite{Bennett1993}, measurement is a crucial operation because it induces the photonic nonlinearities needed for more advanced information processing.  Therefore, the ability to actively switch a quantum state conditioned on such measurements is a key enabling technology. Pertinent examples include multiplexing probabilistic sources~\cite{GrimauPuigibert2017,Kaneda2019,Hiemstra2020}, implementing gates in photonic quantum computers~\cite{Pittman2002,Prevedel2007}, changing bases in active quantum teleportation protocols~\cite{Giacomini2002,Pirandola2015}, switching photons in quantum state memories~\cite{Pittman2002a}, quantum state generation~\cite{Massaro2019,LuizZanin2021,Meyer-scott2022,Hou2023}, quantum state correction~\cite{Donaldson2019}, and quantum metrology~\cite{Sabines-chesterking2017}. 

\begin{figure}
    \centering
    \includegraphics[width=0.9\linewidth]{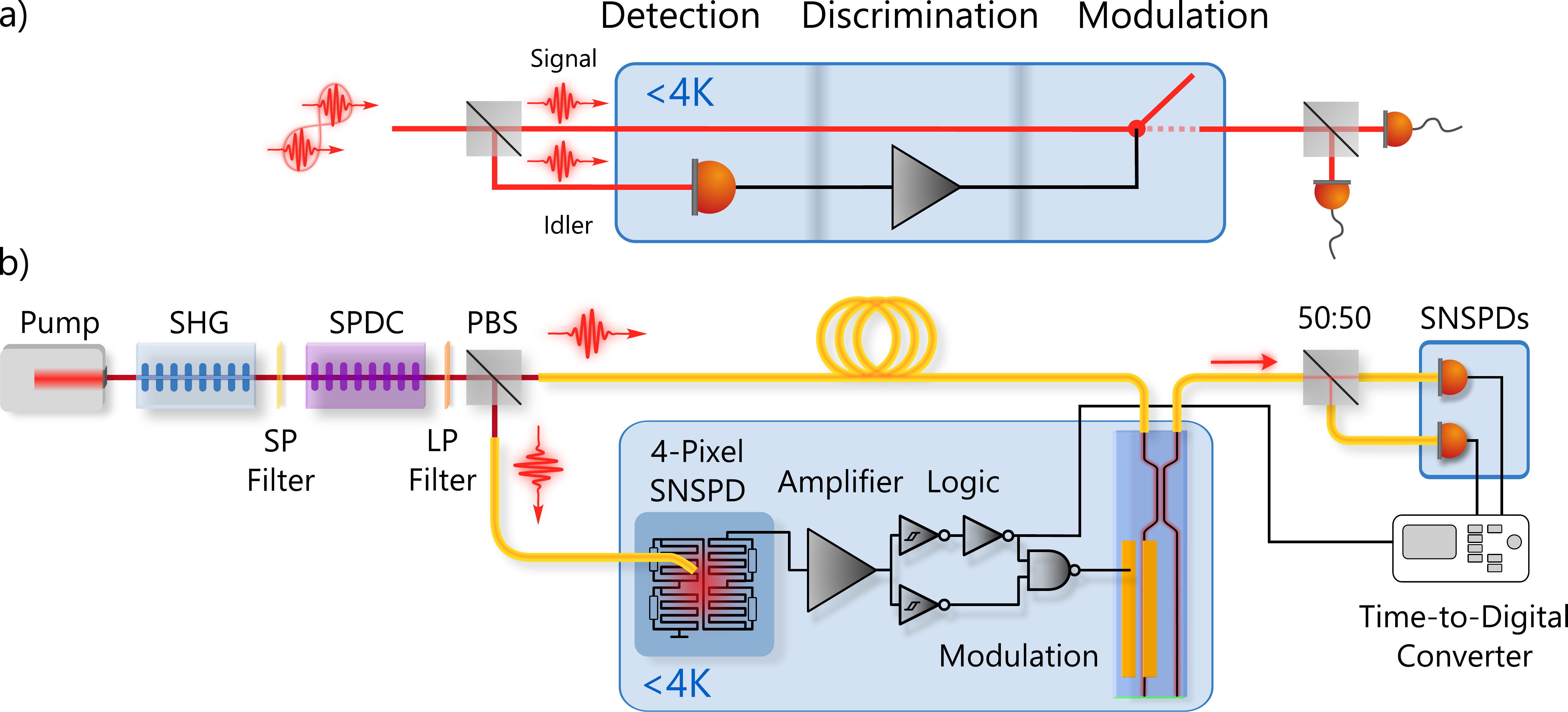}   
    \caption{a) Schematic of a feedforward operation, where the measurement, discrimination, and modulation all take place under cryogenic conditions. A quantum state from spontaneous parametric down-conversion (SPDC) is separated into its signal and idler modes. At cryogenic temperatures, the photon number of the idler mode is measured. Based on a preset photon number outcome, the electro-optic modulator is set to high or low transmission for the signal photons, in order to filter a specific photon-number state in the signal mode. The resulting output signal is analyzed with a Hanbury-Brown-Twiss (HBT) interferometer. b) Detailed layout of the feedforward operation. The quantum states are generated in a type-II SPDC source at \SI{1556}{nm} which is pumped with up-converted light from a Second Harmonic Generation (SHG) source. The pump light is filtered with appropriately chosen short- (SP) and long-pass (LP) filters. After splitting signal and idler modes at a Polarizing Beam Splitter (PBS), the quantum state is actively manipulated at cryogenic temperatures. This relies on measurement of the idler mode by a multipixel SNSPD, followed by co-located amplification, logic processing, and electro-optic modulation.  The HBT interferometer uses Superconducting Nanowire Single-Photon Detectors (SNSPD) (located in a different cryostat). The SNSPD output signals are acquired with a time-to-digital converter in combination with the output signal from the logic circuit.}
    \label{Overview}
\end{figure}

In general, feedforward requires measurement, signal acquisition and processing, followed by optical modulation. The latency between measurement and modulation is a key figure of merit in all these applications. Reducing latency is often crucial, since losses and decoherence scale with delay or storage time. This means minimizing not only the processing time of each component, but also the signal transmission time between each element. To the best of our knowledge, the lowest latency quantum photonic feedforward experiment comprising single-photon measurements, logic operations and modulation is \SI{100}{ns}, performed by Pittman et al.~\cite{Pittman2002} and Kaneda et al.~\cite{Kaneda2019}. These experiments both used Avalanche Photodiodes (APDs) operating at room-temperature, in close proximity to logic electronics and an optical modulator. Other experiments using APDs to drive a modulator directly have achieved switching delays as low as~\SI{20}{ns}, but lacked logic processing~\cite{Giacomini2002,Mikova2012}. Although functional at room-temperature, APDs generally do not meet the efficiency, noise, and response time requirements for advanced quantum photonics in the technologically relevant telecom wavelength regime.

Superconducting Nanowire Single-Photon Detectors (SNSPDs) have addressed these limitations, offering near-unity detection efficiency~\cite{Reddy2020}, low timing jitter~\cite{Korzh2020}, and minimal dark counts~\cite{Chiles2022}. Nevertheless, this is at the cost of operating temperatures below \SI{4.2}{K}~\cite{Natarajan2012c}. Initial implementations of feedforward using SNSPDs~\cite{Massaro2019,Hiemstra2020,LuizZanin2021} have kept the SNSPDs in a cryogenic environment, separate from room temperature photonic elements and processing electronics. Interfaces between these elements introduce transmission latency which must be compensated by optical delays. Currently, latency times using SNSPDs are limited to \SI{512}{ns}~\cite{GrimauPuigibert2017}. Therefore, minimizing the physical distance between such detectors, signal processing electronics and an optical switching device within a cryostat is crucial. 

Interfacing SNSPDs directly with electro-optic modulators has previously been investigated in a cryogenic environment~\cite{DeCea2020,Thiele2023}. These experiments were limited by the voltage mismatch between the output of the SNSPD and the drive voltage of the modulator, resulting in a low extinction ratio. Using intermediate amplifiers not only enables the switching voltages required for full state transformation, but also enables more complex logical circuits.  These amplifiers need to bridge the gap between a \SI{1}{mV} detection signal amplitude and the \SI{100}{mV}-\SI{3}{V} required for cryogenic electro-optic modulators and typical CMOS logic gates. Furthermore, the amplifiers and modulators must maintain a high signal integrity by electronic bandwidth matching and a low thermal dissipation to minimize heating the cryostat. 

In addition to the detection and electronic signal processing, the optical modulation itself must also be performed under cryogenic operating conditions. Recently, a number of devices have demonstrated high-fidelity opto-electronic interconnects in a cryogenic environment~\cite{Youssefi2021,Thiele2022c,Thiele2023,Shen2024,Thiele2024b}. Most of these modulators use the electro-optic coefficient in lithium niobate, since this provides both the high switching speeds and low heat dissipation required for such devices. Indeed, lithium niobate has the added advantage of being an excellent photonic integration platform~\cite{Sharapova2017b,Luo2019}, with capacity for cryogenic sources~\cite{Bartnick2021,Lange2022,Lange2023}, cryogenic modulation~\cite{Yoshida1999a,Thiele2020,Thiele2022,Lomonte2021b,Shen2024} and integrated superconducting detectors~\cite{Tanner2012,Hoepker2019,Lomonte2021b,Hoepker2021b}. Competing cryogenic technologies for electro-optic modulators have been demonstrated with semiconductor plasma dispersion~\cite{Gehl2016a,DeCea2020}, Micro-Electro-Mechanical Systems (MEMS)~\cite{Gyger2021,Beutel2022}, or thermal phase-shifters~\cite{Alexander2024}. However, these modulators either suffer from higher optical losses, increased heat load or limited speed which limits their application for cryogenic quantum applications. Alongside the electro-optic requirements, these modulators must be reliably coupled to minimize transmission losses. While active cryogenic fiber alignment can be used to test individual devices~\cite{Wolff2021}, the long-term goal is robust packaging which remains efficiently coupled during thermal transitions~\cite{Wasserman2022,Lin2023,Zeng2023}.

In this paper, we perform a feedforward operation on correlated photon pairs with a tightly packaged SNSPD, amplifier and electro-optic modulator co-located in a cryostat, as shown in Fig.~\ref{Overview}~a). Correlated photons are generated by Spontaneous Parametric-Down Conversion (SPDC) at room temperature and split into idler and signal paths by their polarization. The idler photon is detected by a commercial four-pixel SNSPD. Depending on the number of pixels which fire, an amplitude multiplexed signal is generated, providing quasi-photon number resolution (PNR) up to four photons~\cite{Tiedau2020,Schapeler2020}. The cryogenic logic circuit amplifies the detection signal and discriminates the signal's amplitude. This generates a modulation signal that either transmits or rejects incident signal photons in the modulator.  For this we use an integrated electro-optic phase modulator in titanium in-diffused lithium niobate (Ti:LN) at \SI{3.4}{K}~\cite{Thiele2020,Thiele2022,Thiele2023}. We analyze the transmitted single photons using a Hanbury Brown and Twiss (HBT) interferometer with SNSPDs (located in a second cryostat) to measure single-photon correlations. The cryogenic quantum photonic circuit is shown schematically in Fig.~\ref{Overview}~b).

We demonstrate a proof-of-principle operation of the feedforward circuit by selectively transmitting signal photons based on the idler's photon number. Multi-photon events generated by the SPDC process can be suppressed or enhanced through PNR selection. This actively changes the photon statistics of the signal mode, which we identify by measuring the $g^{(2)}(0)$ function. Our circuit demonstrates the efficacy of combining superconducting detectors, semiconductor-based signal processing, and modulation in a cryogenic environment. 

\section{Results}\label{sec:results}
\begin{figure}
    \centering
    \includegraphics[width=0.68\linewidth]{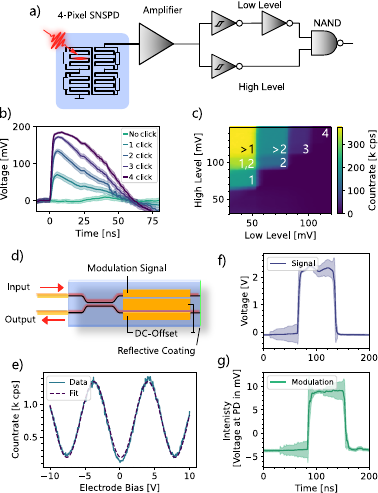}
    \caption{a) Layout of the 4-pixel SNSPD, amplifier and discriminator circuit. The SNSPD is located at the \SI{0.9}{K} stage of the cryostat, while the amplifier and logic are mounted to the \SI{4}{K} stage. b) Averaged traces from the 4-pixel SNSPD selected by their amplitude after the first stage amplifier. c) Count rate of the 4-pixel SNSPD and logic circuit based on the detection signals. The low- and high level threshold for the discriminator circuit are swept. Threshold ranges of similar count rates are identified as the photon number outputs. d) Layout of the electro-optic modulator in a Michelson-interferometer configuration. The 50:50 beamsplitter, phase modulator and reflective end-facet are cointegrated in a titanium in-diffused lithium niobate waveguide chip. e) Optical modulation during a voltage sweep when illuminated by the signal mode from the SPDC source. f) Time-trace of the electrical modulator actuation signal after a detection event (acquired with an oscilloscope at room temperature with a \SI{50}{\ohm} load). g) Optical response of the cryogenic modulator given a modulation signal, acquired at room temperature by a photodiode. In b), f) and g) the signals are averaged and the bands show the standard deviation in the respective signals.}
    \label{SumResults}
\end{figure}
\subsection{Component performance}

\textbf{Detection:} 
The first step of our feedforward operation is to measure the number of photons in the idler mode of the SPDC source with a four-pixel SNSPD, as shown in Fig.~\ref{SumResults}~a). We use a commercial multipixel SNSPD, comprising four separate meanders, each with its own shunt resistor. When at least one photon is absorbed in a meander, the bias current is redirected to the shunt resistor, resulting in a voltage signal in the sub millivolt range. Chaining the four pixels in series enables us to add up their signals, such that when multiple pixels are simultaneously triggered a larger signal is produced, as shown in Fig.~\ref{SumResults}~b). The amplitude multiplexing provides a quasi-photon-number resolution encoded in the signal amplitude~\cite{Tiedau2020}, which can be discriminated by the readout electronics. More details on the detector characterization and PNR of this detector can be found in Schapeler et al. \cite{Schapeler2020} and the Supplementary Material.

\textbf{Discrimination:} The detection signal is then transmitted to the cryogenic logic circuit for further amplification, discrimination, and electronic pulse shaping. Initially, the detection pulses are amplified with a chain of common emitter amplifiers with a total gain of approximately \SI{60}{dB}, increasing the signal from the sub-mV range to about \SI{100}{mV}, see Fig.~\ref{SumResults}~b). Next, the amplified signal is discriminated by the peak amplitude. To do so, the signal is split into two Schmitt-trigger circuits with an inverting and non-inverting arm as seen in Fig.~\ref{SumResults}~a). When a preset threshold is surpassed, either trigger-circuit generates a CMOS-compatible trigger signal. For a further comparison, the trigger signals are sent to a NAND-gate. By setting the threshold level on the Schmitt-trigger, we can select upper and lower boundaries in the signal amplitude to generate a modulation signal, as shown in Fig.~\ref{SumResults}~c). A sweep of the high and low level threshold shows the change in the count rate while illuminating the SNSPD with an attenuated pulsed coherent state. A clear plateau in the count rates can be observed which corresponds to signals only in a given amplitude range. As a result, we can freely choose to generate modulation signals for e.g. exactly one, exactly two or more than two photons. The generated modulation signal switches between \SI{0}{V} and \SI{3.6}{V} supply voltage, has a rise time below \SI{5}{ns} and a pulse length of \SI{80}{ns}, as shown in Fig.~\ref{SumResults}~f). The logic circuit achieves an electronic delay between the SNSPD’s input and the modulation output of approximate \SI{23(3)}{ns}. The majority of this delay with \SI{20(2)}{ns} is due to the trigger circuits and logic gates. We based our logic circuit on SiGe-BiCMOS components which an enable a heat dissipation as low as \SI{35}{mW} at \SI{3.4}{K} (with more details in section~\ref{sec:Meth}) and Supplementary Material).

\textbf{Modulation:} Our aim is to use the modulation signal generated by idler events to actively manipulate the signal photons. We designed and fabricated an integrated Michelson interferometer in a titanium in-diffused lithium niobate (Ti:LN) chip as shown in Fig.~\ref{SumResults}~d). The signal mode is coupled into and out of the photonic chip using dual-core single-mode fiber pigtail glued to one end-facet. Once in the chip, the signal mode is incident on a 50:50 beam splitter, whose outputs comprise the two paths of the Michelson interferometer. Electrodes on each path of the interferometer form electro-optic phase shifters. The light is then reflected back at the opposite end-facet by a reflective coating. After propagating back, the light interferes at the beamsplitter. Depending on the applied voltage on the phase shifters, light is either emitted through the input or output fibers. This combination of Michelson interferometer and modulator thus acts as a controllable mirror. Moreover, this configuration doubles the effective length of the electrodes, which reduces the voltage $V_{\pi}$ required to switch between the in- and output ports of the device to \SI{3.82(0.05)}{V}, as can be seen in Fig.~\ref{SumResults}~e). The electrodes are placed in a signal-ground-signal configuration with one side as the signal input and the other as a voltage off-set. At \SI{1556}{nm}, we achieve an optical modulation depth of \SI{10.2(0.2)}{dB}, given an \SI{8}{nm} optical bandwidth in the full width at half maximum (FWHM) of the PDC photons, and a \SI{3.6}{V} modulation signal. The fiber connection achieves a transmission of 8\% from fiber-to-fiber at cryogenic temperatures, which includes interface losses and fiber-waveguide mode mismatch, all on-chip losses, losses at the reflective coating, and the visibility of the Michelson interferometer. Further information on the cryogenic electro-optic modulator’s electronic bandwidth, transmission, fiber access, and wavelength-dependent transmission is provided in the supplementary material.

\subsection{Real-time manipulation of quantum light}

\begin{figure}
    \centering
    \includegraphics[width=0.85\linewidth]{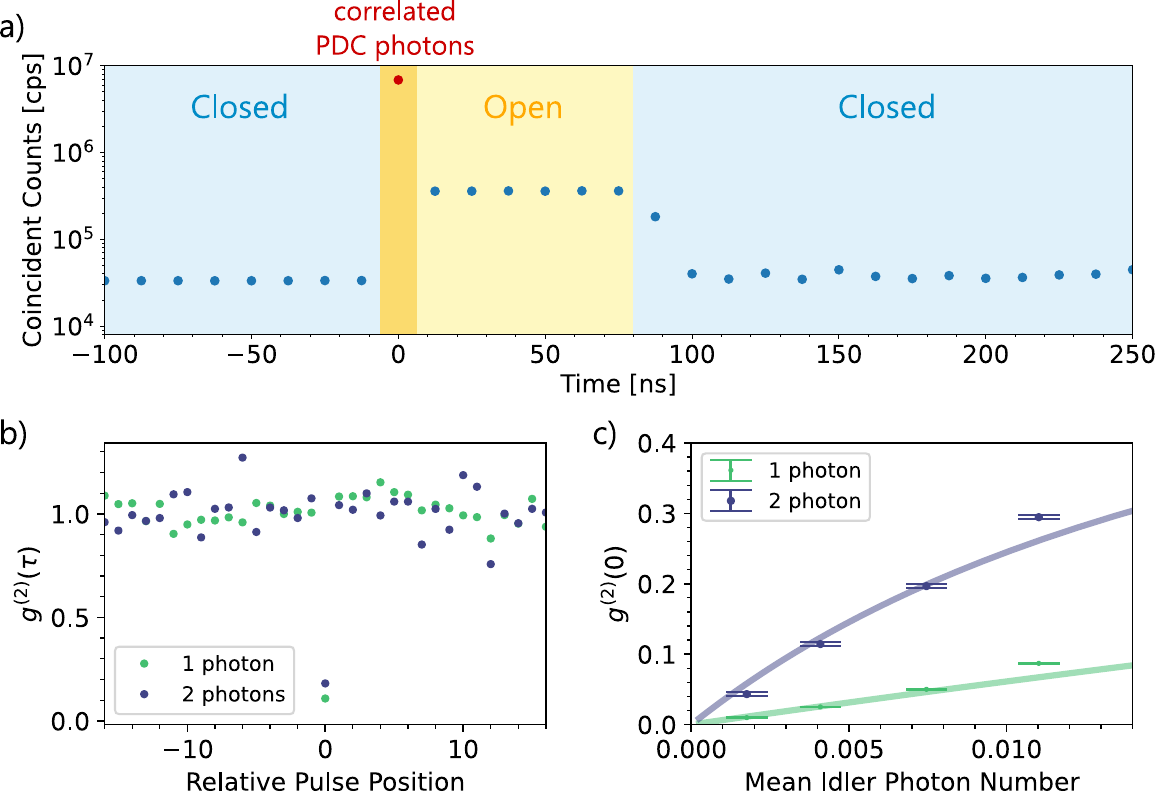}
    \caption{a) The coincidence count rate between the modulator switching signal and a photon detection event on one of the detectors in the HBT-interferometer. The mean photon number for this measurement was 0.0075 photons per pulse. Colored regions indicate the transmission state of the modulator: “open'' (yellow) corresponds to high transmission, “closed'' (blue) to low transmission. Correlated photons are colored red. The “open''-region is equivalent to a high modulation voltage in~\ref{SumResults}~g). b) The second order correlation function $g^{(2)}(\tau)$ when switching the modulator while triggering on a one photon (green) or two photons (blue) incident at the idler detector. c) The second order correlation function at zero time delay $g^{(2)}(0)$ for different mean photon numbers in the idler mode. The simulated $g^{(2)}(0)$, accounting for losses and cross-talk in the herald detector, is indicated by the lines.} 
    \label{fig:g2}
\end{figure}

\textbf{Circuit Operation:} 
Our first goal is to successfully demonstrate switching a single photon, conditional on the detection of another photon. To do so, we measure coincidence count-rates between the idler measurement and the signal mode transmitted through the modulator and detected with a separate SNSPD.  We measure coincidences within a time-bin of \SI{250}{ps}; this results in a series of peaks with \SI{12.5}{ns} spacing due to the laser repetition rate. After integrating over the peaks, we obtain the data shown in Fig.~\ref{fig:g2}~a). When an idler photon is measured by any one of the four pixels, it triggers the modulator to switch to the “open" position for $\sim$\SI{80}{ns}. This allows transmission of both the correlated signal photons, followed by uncorrelated SPDC photons generated from later pulses, through the modulator. The modulator then returns to the “closed" state. 

From figure~\ref{fig:g2}~a), we can clearly see the modulator switches at the appropriate time and lets through the correlated photons from the down-conversion. The relative ratio of the subsequent uncorrelated peaks in the “open" position to the peaks when the modulator is “closed" shows that the feedforward circuit suppresses uncorrelated events by \SI{10.2(0.2)}{dB}. 

\textbf{Feedforward control of photon statistics:} Using the multipixel quasi-photon-number-resolving detector, we can investigate how modulation based on photon number measurements can influence the photon number statistics of the signal photons. To do so, we use the second order correlation function $g^{(2)}(\tau)$, which we measure using a HBT-interferometer as shown in Fig.~\ref{Overview}~b).  In this type of measurement the second-order correlation function $g^{(2)}(\tau)$ can be approximated in the low photon number regime by $g^{(2)}(\tau)\approx \frac{C_{ab}(\tau) f_\textrm{rep} }{C_{a} C_{b}}$ , where $C_{i}$ is the count rate at detector $i\in \{a,b\}$  and $C_{ab}(\tau)$ is the coincidence count rate between the detectors at a given time delay $\tau$, at a laser repetition rate $f_\textrm{rep}$ ~\cite{Bashkansky2014}. In this way, we can measure the $g^{(2)}(\tau)$ given modulation conditioned on different detected photon numbers in the idler mode. 

We investigate cases of switching the modulator following single- and two-photon detection events at the multipixel detector. For an ideal source and detector, the proportion of two-photon events to single-photon events should follow a thermal distribution. Nevertheless, imperfections such as a multi-thermal source~\cite{Avenhaus2008} and cross-talk in the detector~\cite{Schapeler2020} can change the statistics. At the highest pump power, the single- and two-photon count rates of the idler modes are 825000 counts per second and 56000 counts per second, respectively.

The $g^{(2)}(\tau)$ of the modulated signal mode, conditional on one or two photons measured in the idler mode, is shown in Fig.~\ref{fig:g2}~b). For measurement events in the same pulse, there is a clear dip below 1. As expected, the $g^{(2)}(0)$  is close to 0 for a single photon. For the two photon events it is slightly higher, nevertheless we expect a $g^{(2)}(0)=0.5$. This discrepancy is due to cross-talk in our multipixel heralding detector. Cross-talk means single photons in the idler mode cause two pixels to fire, and are thus regarded as a two-photon event when heralding, when in fact there are only single photons in signal mode measured by the HBT setup. In our commercial multipixel detector, the cross-talk probability is around 2.5\%. Based on this probability, we can estimate that 63\% of the observed two-photon clicks are genuine two-photon events, and 37\% are caused by single photons and cross-talk. This is further demonstrated in Fig.~\ref{fig:g2}~c), where we measure the $g^{(2)}(0)$  as we increase the pump power. Here, the two-photon events begin to dominate as the pump power increases, and the $g^{(2)}(0)$ tends towards the expected value of 0.5. The single-photon $g^{(2)}(0)$ also increases slightly with pump power, due to limited detection efficiency, which manifests as higher photon numbers appearing as single photons when measured. Due to counting in coincidence mode we can neglect dark counts, given a raw dark-count rate of around 100 counts per second. A more thorough formalism of the detector including imperfections such as loss, limits of photon-number resolution and cross-talk, is defined by a Positive Operator Value Measure (POVM) and is provided in the Supplementary Material~\ref{sec:sup}.

\section{Discussion}\label{sec:Disc}
All photonic feedforward operations rely on measurements at the single-photon level, followed by discrimination and active manipulation of a quantum state. We have shown that this is possible using SNSPDs, electronic amplification, logic and electro-optic switching, all co-located within a cryostat at \SI{4}{K}. This paves the way for more complex circuits and photonic quantum state control with low latency.

Our specific feedforward measurement signal leveraged the discrimination of different photon numbers. This type of measurement will be relevant not only in multiplexing probabilistic quantum light sources, but also multi-photon error identification and correction. While we can clearly demonstrate switching based on measurements of different photon numbers, we were limited by the effects of cross-talk in our commercial multipixel detector. This effect can be reduced with recent advances in multipixel devices~\cite{Stasi2024} and intrinsic photon number resolution in SNSPDs based on rise-time analysis~\cite{Cahall2017,Sauer2023,Schapeler2024,Kong2024}, although readout electronics capable of this discrimination at cryogenic temperatures have yet to be demonstrated.

Our demonstration of the colocation of photonic and electronic components under cryogenic conditions brings significant scalability benefits. Multiple logic operations can be linked \emph{in situ} to implement more complex feedforward operations at high speed. Increasing the size and complexity of feedforward circuits will inevitably add heat dissipation. Nevertheless, our current circuit dissipates \SI{35}{mW} of heat during operation; by comparison, commercial cryostats offer up to the order of \SI{1}{W} cooling power at \SI{4}{K}. Furthermore, \emph{in situ} operation reduces the number of high-speed interconnects between room temperature and the cold stage: each such interconnect (typically coax-lines), dissipate heat in the \SIrange{10}{100}{mW} range~\cite{Krinner2019,Thiele2024b}.  

\begin{figure}
    \centering
    \includegraphics[width=0.5\linewidth]{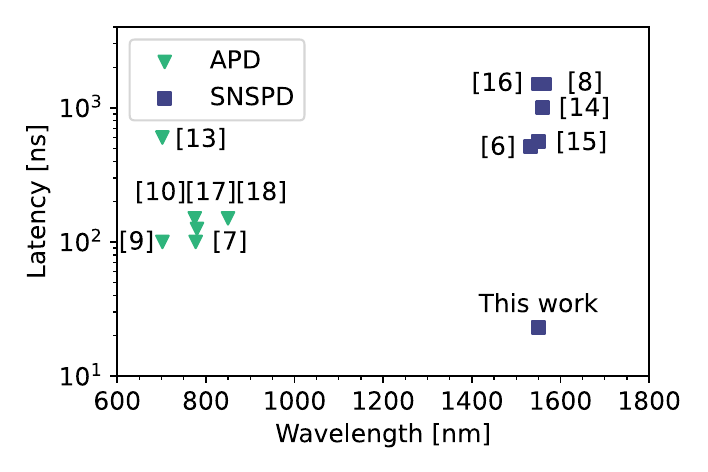}
    \caption{Comparison of the latency and operating wavelength in previous quantum feedforward operations. The latency includes the idler photon detection, amplification, logic processing and optical modulation. The feedforward operations with APDs are indicated by green triangles and with SNSPDs by blue squares.}
    \label{fig:LitComp}
\end{figure}

In our setup, we achieve an electronic delay of the logic circuit of \SI{23(3)}{ns}, which is nearly an order of magnitude faster than the  state of the art, as shown in Fig.~\ref{fig:LitComp}. In contrast to the earlier experiments, we circumvent added transmission delays between the detector and logic processing in coax-lines (\SI{5}{ns/m}). Furthermore, realizing the amplifier and discriminator circuits with an integrated SiGe-BiCMOS circuit would reduce latency to a few nanoseconds and lower heat load.  This latency determines the required storage time of the remaining state, which in turn affects the storage losses. Optical delay lines of the order of this latency can be realized using integrated photonic waveguides, with losses as low as \SI{0.08}{dB/m} in silica ~\cite{Lee2012}. Alternatively, low-loss single-mode fibers could be used, provided effective low-loss coupling to the waveguide is achieved. Our Ti:LN device offers robust cryogenic optical coupling in a packaged module, without requiring additional optics for mode conversion to single-mode fibers at telecom wavelengths. Improvements to optical packaging and mode matching in extreme environments is ongoing~\cite{Wasserman2022,Lin2023,Zeng2023}.

By using phase modulators in a Michelson configuration, we were able to switch the full \SI{8}{nm} bandwidth photons from the SPDC source with a modulation depth of \SI{10.2(0.2)}{dB}. The bandwidth of the optical pulses which can be modulated is ultimately constrained by the interferometer's free-spectral range. This tends to be large for Michelson or Mach-Zehnder configurations, and would be further improved by reducing the device dimensions. Miniaturization of the optics is also a clear path to increasing complexity. This could be achieved in other electro-optic waveguide platforms such as thin-film lithium niobate (TFLN)~\cite{Zhu2021,Hu2024} or lithium tantalate~\cite{Wang2024}. Indeed, TFLN has already proven to offer switching voltages compatible with CMOS logic under cryogenic conditions~\cite{Shen2024}.

We have shown a fully cryogenic feedforward switch based on a photon-number-resolved measurement of a photonic mode. We used this device to actively herald single- and two-photon states from a SPDC source. To do so, it was necessary to combine both conventional and quantum integrated photonics with superconducting and semiconducting electronics in a cryogenic environment. By converting single-photon measurements to electronic logic \emph{in situ}, we open up the possibility to perform high-speed control over photonic quantum systems, which is a cornerstone of quantum information processing and communication. 

\section{Methods} \label{sec:Meth}
\textbf{Electrical circuit}
The electrical circuit used to perform the feedforward operation is made up of discrete SiGe Heterojunction-Bipolar-Transistors (HBTs) and CMOS logic gates. For passive electrical elements we use NiCr-thinfilm resistors and ceramic capacitors. The components are soldered on a printed circuit board and connected using coaxial cables inside the cryostat.
The first stage of the electronic circuit is a set of amplifiers to boost the weak signal of the SNSPD array in the order of a millivolt to the order of \SI{100}{mV} for further processing. For this we employ a low noise amplifier design with a stabilizing feedback loop from the collector to the transistor base. To reach a sufficiently high voltage level three of these amplifiers are used in series.

In order to select photon numbers states from the detector output we need to be able to trigger on different signal amplitudes. This is achieved using a differential amplifier with a capacitively coupled positive feedback loop. The capacitive feedback ensures the trigger will reset itself after a set on-time. The trigger levels can be set using one of the inputs of the differential amplifier.
Selecting individual photon numbers requires the use of two triggers with an additional inverter and NAND gate to select for example “more than 1 photon AND NOT 2 photons”. With the digital signal generated by the NAND gate the modulator can be switched. For this we use a CMOS buffer with a larger output current to ensure fast turn on times. The full circuit diagrams and a list of the used components can be seen in the Supplementary Material.

The total processing delay can be estimated as follows. From the roughly \SI{30}{cm} of coaxial cabling in the setup we can estimate a delay of \SI{1.5}{ns}, assuming a transmission speed of \SI{0.66}{c} as is typical for coaxial cables. The modulator can be estimated to have a rise time of \SI{1.5}{ns} based on its cryogenic bandwidth of \SI{230}{MHz}. The amplifiers are not considered for the delay as high frequency amplifiers typically have group delays in the picosecond range. By far the largest contribution comes from the trigger and logic board with \SI{20}{ns}. This delay was directly measured by using an artificially generated SNSPD click signal at the input of the trigger and comparing the delay to the same signal but omitting the circuit board. 

\textbf{Simulation}
To validate the observed $g^{(2)}(0)$ values we simulate the effect of the feedforward circuit on the observed photon statistics. To this end, we compute the probability for a given number of created pairs to be detected as the photon number chosen in the discrimination circuit in the idler mode. For this we take coupling loss as well as detector non-ideality into account. Based on this detection probability and the photon-pair-number distribution created by the SPDC source, we compute the photon-number distribution in the signal mode. We consider only detection events in which heralding events have taken place. From the resulting photon-number distribution we can compute the expected $g^{(2)}(0)$ values. Further details are provided in the supplementary materials.

\textbf{Source}
We use correlated photons generated in the signal and idler modes of a Spontaneous Parametric-Down Conversion (SPDC) process as our source of quantum light. To generate these photons, we start with sub-picosecond laser pulses at \SI{1556}{nm}, which are up-converted to \SI{778}{nm} in a second-harmonic generation process in a bulk MgO-doped periodically poled lithium niobate crystal. These up-converted photons then act as the pump in the SPDC process to produce correlated photon pairs at \SI{1556}{nm}, in a Periodically Poled Potassium Titanyl Phosphate (ppKTP) waveguide. Both the SHG and SPDC processes are conducted at room temperature with appropriate filtering of pump light using short and long pass filters. We use a type-II SPDC process to allow us to deterministically split the signal and idler modes by their orthogonal polarization into two paths with a polarizing beam splitter. The idler channel is transmitted to the multipixel superconducting detector, while the signal channel is transmitted to the input of the modulator. Further details on the source are provided in the supplementary information.

\backmatter

\bmhead{Supplementary information}~\label{sec:sup}
Supplementary material is additionally provided.
All datasets used in this publication in the main manuscript and the supplementary material are openly available (\url{https://doi.org/10.5281/zenodo.13753477}).

\bmhead{Acknowledgements}
This work was supported by the Bundesministerium für Bildung und Forschung (BMBF) under the QPIC-1 project (Grant No. 13N15856). Funded by the European Union (ERC, QuESADILLA, 101042399). Views and opinions expressed are however those of the author(s) only and do not necessarily reflect those of the European Union or the European Research Council Executive Agency. Neither the European Union nor the granting authority can be held responsible for them. We thank Timon Schapeler for setting up the POVM-matrix of the four-pixel SNSPD.

\bibliography{sn-bibliography}


\begin{thebibliography}{68}
\ifx \bisbn   \undefined \def \bisbn  #1{ISBN #1}\fi
\ifx \binits  \undefined \def \binits#1{#1}\fi
\ifx \bauthor  \undefined \def \bauthor#1{#1}\fi
\ifx \batitle  \undefined \def \batitle#1{#1}\fi
\ifx \bjtitle  \undefined \def \bjtitle#1{#1}\fi
\ifx \bvolume  \undefined \def \bvolume#1{\textbf{#1}}\fi
\ifx \byear  \undefined \def \byear#1{#1}\fi
\ifx \bissue  \undefined \def \bissue#1{#1}\fi
\ifx \bfpage  \undefined \def \bfpage#1{#1}\fi
\ifx \blpage  \undefined \def \blpage #1{#1}\fi
\ifx \burl  \undefined \def \burl#1{\textsf{#1}}\fi
\ifx \doiurl  \undefined \def \doiurl#1{\url{https://doi.org/#1}}\fi
\ifx \betal  \undefined \def \betal{\textit{et al.}}\fi
\ifx \binstitute  \undefined \def \binstitute#1{#1}\fi
\ifx \binstitutionaled  \undefined \def \binstitutionaled#1{#1}\fi
\ifx \bctitle  \undefined \def \bctitle#1{#1}\fi
\ifx \beditor  \undefined \def \beditor#1{#1}\fi
\ifx \bpublisher  \undefined \def \bpublisher#1{#1}\fi
\ifx \bbtitle  \undefined \def \bbtitle#1{#1}\fi
\ifx \bedition  \undefined \def \bedition#1{#1}\fi
\ifx \bseriesno  \undefined \def \bseriesno#1{#1}\fi
\ifx \blocation  \undefined \def \blocation#1{#1}\fi
\ifx \bsertitle  \undefined \def \bsertitle#1{#1}\fi
\ifx \bsnm \undefined \def \bsnm#1{#1}\fi
\ifx \bsuffix \undefined \def \bsuffix#1{#1}\fi
\ifx \bparticle \undefined \def \bparticle#1{#1}\fi
\ifx \barticle \undefined \def \barticle#1{#1}\fi
\bibcommenthead
\ifx \bconfdate \undefined \def \bconfdate #1{#1}\fi
\ifx \botherref \undefined \def \botherref #1{#1}\fi
\ifx \url \undefined \def \url#1{\textsf{#1}}\fi
\ifx \bchapter \undefined \def \bchapter#1{#1}\fi
\ifx \bbook \undefined \def \bbook#1{#1}\fi
\ifx \bcomment \undefined \def \bcomment#1{#1}\fi
\ifx \oauthor \undefined \def \oauthor#1{#1}\fi
\ifx \citeauthoryear \undefined \def \citeauthoryear#1{#1}\fi
\ifx \endbibitem  \undefined \def \endbibitem {}\fi
\ifx \bconflocation  \undefined \def \bconflocation#1{#1}\fi
\ifx \arxivurl  \undefined \def \arxivurl#1{\textsf{#1}}\fi
\csname PreBibitemsHook\endcsname

\bibitem[\protect\citeauthoryear{Bennett and Brassard}{1984}]{BB84}
\begin{bchapter}
\bauthor{\bsnm{Bennett}, \binits{C.H.}},
\bauthor{\bsnm{Brassard}, \binits{G.}}:
\bctitle{{Quantum cryptography: Public key distribution and coin tossing}}.
In: \bbtitle{International Conference on Computers, Systems {\&} Signal Processing},
\bconflocation{Bangalore, India},
pp. \bfpage{175}--\blpage{179}
(\byear{1984})
\end{bchapter}
\endbibitem

\bibitem[\protect\citeauthoryear{Aaronson and Arkhipov}{2011}]{Aaronson2011}
\begin{bchapter}
\bauthor{\bsnm{Aaronson}, \binits{S.}},
\bauthor{\bsnm{Arkhipov}, \binits{A.}}:
\bctitle{{The computational complexity of linear optics}}.
In: \bbtitle{Proceedings of the Forty-third Annual ACM Symposium on Theory of Computing},
pp. \bfpage{333}--\blpage{342}.
\bpublisher{ACM},
\blocation{New York, NY, USA}
(\byear{2011}).
\doiurl{10.1145/1993636.1993682} .
\burl{https://dl.acm.org/doi/10.1145/1993636.1993682}
\end{bchapter}
\endbibitem

\bibitem[\protect\citeauthoryear{Hamilton et~al.}{2017}]{Hamilton2017}
\begin{barticle}
\bauthor{\bsnm{Hamilton}, \binits{C.S.}},
\bauthor{\bsnm{Kruse}, \binits{R.}},
\bauthor{\bsnm{Sansoni}, \binits{L.}},
\bauthor{\bsnm{Barkhofen}, \binits{S.}},
\bauthor{\bsnm{Silberhorn}, \binits{C.}},
\bauthor{\bsnm{Jex}, \binits{I.}}:
\batitle{{Gaussian Boson Sampling}}.
\bjtitle{Physical Review Letters}
\bvolume{119}(\bissue{17}),
\bfpage{170501}
(\byear{2017})
\doiurl{10.1103/PhysRevLett.119.170501}
{\href{https://arxiv.org/abs/1612.01199}{{arXiv:1612.01199}}}
\end{barticle}
\endbibitem

\bibitem[\protect\citeauthoryear{Zhong et~al.}{2020}]{Zhong2020}
\begin{barticle}
\bauthor{\bsnm{Zhong}, \binits{H.-s.}},
\bauthor{\bsnm{Wang}, \binits{H.}},
\bauthor{\bsnm{Deng}, \binits{Y.-h.}},
\bauthor{\bsnm{Chen}, \binits{M.-c.}},
\bauthor{\bsnm{Peng}, \binits{L.-c.}},
\bauthor{\bsnm{Luo}, \binits{Y.-h.}},
\bauthor{\bsnm{Qin}, \binits{J.}},
\bauthor{\bsnm{Wu}, \binits{D.}},
\bauthor{\bsnm{Ding}, \binits{X.}},
\bauthor{\bsnm{Hu}, \binits{Y.}},
\bauthor{\bsnm{Hu}, \binits{P.}},
\bauthor{\bsnm{Yang}, \binits{X.-Y.}},
\bauthor{\bsnm{Zhang}, \binits{W.-j.}},
\bauthor{\bsnm{Li}, \binits{H.}},
\bauthor{\bsnm{Li}, \binits{Y.}},
\bauthor{\bsnm{Jiang}, \binits{X.}},
\bauthor{\bsnm{Gan}, \binits{L.}},
\bauthor{\bsnm{Yang}, \binits{G.}},
\bauthor{\bsnm{You}, \binits{L.}},
\bauthor{\bsnm{Wang}, \binits{Z.}},
\bauthor{\bsnm{Li}, \binits{L.}},
\bauthor{\bsnm{Liu}, \binits{N.-l.}},
\bauthor{\bsnm{Lu}, \binits{C.-y.}},
\bauthor{\bsnm{Pan}, \binits{J.-w.}}:
\batitle{{Quantum computational advantage using photons}}.
\bjtitle{Science}
\bvolume{370}(\bissue{6523}),
\bfpage{1460}--\blpage{1463}
(\byear{2020})
\doiurl{10.1126/science.abe8770}
\end{barticle}
\endbibitem

\bibitem[\protect\citeauthoryear{Madsen et~al.}{2022}]{Madsen2022}
\begin{barticle}
\bauthor{\bsnm{Madsen}, \binits{L.S.}},
\bauthor{\bsnm{Laudenbach}, \binits{F.}},
\bauthor{\bsnm{Askarani}, \binits{M.F.}},
\bauthor{\bsnm{Rortais}, \binits{F.}},
\bauthor{\bsnm{Vincent}, \binits{T.}},
\bauthor{\bsnm{Bulmer}, \binits{J.F.F.}},
\bauthor{\bsnm{Miatto}, \binits{F.M.}},
\bauthor{\bsnm{Neuhaus}, \binits{L.}},
\bauthor{\bsnm{Helt}, \binits{L.G.}},
\bauthor{\bsnm{Collins}, \binits{M.J.}},
\bauthor{\bsnm{Lita}, \binits{A.E.}},
\bauthor{\bsnm{Gerrits}, \binits{T.}},
\bauthor{\bsnm{Nam}, \binits{S.W.}},
\bauthor{\bsnm{Vaidya}, \binits{V.D.}},
\bauthor{\bsnm{Menotti}, \binits{M.}},
\bauthor{\bsnm{Dhand}, \binits{I.}},
\bauthor{\bsnm{Vernon}, \binits{Z.}},
\bauthor{\bsnm{Quesada}, \binits{N.}},
\bauthor{\bsnm{Lavoie}, \binits{J.}}:
\batitle{{Quantum computational advantage with a programmable photonic processor}}.
\bjtitle{Nature}
\bvolume{606}(\bissue{7912}),
\bfpage{75}--\blpage{81}
(\byear{2022})
\doiurl{10.1038/s41586-022-04725-x}
\end{barticle}
\endbibitem

\bibitem[\protect\citeauthoryear{Deng et~al.}{2023}]{Deng2023}
\begin{barticle}
\bauthor{\bsnm{Deng}, \binits{Y.H.}},
\bauthor{\bsnm{Gu}, \binits{Y.C.}},
\bauthor{\bsnm{Liu}, \binits{H.L.}},
\bauthor{\bsnm{Gong}, \binits{S.Q.}},
\bauthor{\bsnm{Su}, \binits{H.}},
\bauthor{\bsnm{Zhang}, \binits{Z.J.}},
\bauthor{\bsnm{Tang}, \binits{H.Y.}},
\bauthor{\bsnm{Jia}, \binits{M.H.}},
\bauthor{\bsnm{Xu}, \binits{J.M.}},
\bauthor{\bsnm{Chen}, \binits{M.C.}},
\bauthor{\bsnm{Qin}, \binits{J.}},
\bauthor{\bsnm{Peng}, \binits{L.C.}},
\bauthor{\bsnm{Yan}, \binits{J.}},
\bauthor{\bsnm{Hu}, \binits{Y.}},
\bauthor{\bsnm{Huang}, \binits{J.}},
\bauthor{\bsnm{Li}, \binits{H.}},
\bauthor{\bsnm{Li}, \binits{Y.}},
\bauthor{\bsnm{Chen}, \binits{Y.}},
\bauthor{\bsnm{Jiang}, \binits{X.}},
\bauthor{\bsnm{Gan}, \binits{L.}},
\bauthor{\bsnm{Yang}, \binits{G.}},
\bauthor{\bsnm{You}, \binits{L.}},
\bauthor{\bsnm{Li}, \binits{L.}},
\bauthor{\bsnm{Zhong}, \binits{H.S.}},
\bauthor{\bsnm{Wang}, \binits{H.}},
\bauthor{\bsnm{Liu}, \binits{N.L.}},
\bauthor{\bsnm{Renema}, \binits{J.J.}},
\bauthor{\bsnm{Lu}, \binits{C.Y.}},
\bauthor{\bsnm{Pan}, \binits{J.W.}}:
\batitle{{Gaussian Boson Sampling with Pseudo-Photon-Number-Resolving Detectors and Quantum Computational Advantage}}.
\bjtitle{Physical Review Letters}
\bvolume{131}(\bissue{15}),
\bfpage{150601}
(\byear{2023})
\doiurl{10.1103/PhysRevLett.131.150601}
{\href{https://arxiv.org/abs/2304.12240}{{2304.12240}}}
\end{barticle}
\endbibitem

\bibitem[\protect\citeauthoryear{Knill et~al.}{2001}]{Knill2001}
\begin{barticle}
\bauthor{\bsnm{Knill}, \binits{E.}},
\bauthor{\bsnm{Laflamme}, \binits{R.}},
\bauthor{\bsnm{Milburn}, \binits{G.J.}}:
\batitle{{A scheme for efficient quantum computation with linear optics}}.
\bjtitle{Nature}
\bvolume{409}(\bissue{6816}),
\bfpage{46}--\blpage{52}
(\byear{2001})
\doiurl{10.1038/35051009}
\end{barticle}
\endbibitem

\bibitem[\protect\citeauthoryear{Raussendorf and Briegel}{2001}]{Raussendorf2001}
\begin{barticle}
\bauthor{\bsnm{Raussendorf}, \binits{R.}},
\bauthor{\bsnm{Briegel}, \binits{H.J.}}:
\batitle{{A one-way quantum computer}}.
\bjtitle{Physical Review Letters}
\bvolume{86}(\bissue{22}),
\bfpage{5188}--\blpage{5191}
(\byear{2001})
\doiurl{10.1103/PhysRevLett.86.5188}
\end{barticle}
\endbibitem

\bibitem[\protect\citeauthoryear{Bennett et~al.}{1993}]{Bennett1993}
\begin{barticle}
\bauthor{\bsnm{Bennett}, \binits{C.H.}},
\bauthor{\bsnm{Brassard}, \binits{G.}},
\bauthor{\bsnm{Cr{\'{e}}peau}, \binits{C.}},
\bauthor{\bsnm{Jozsa}, \binits{R.}},
\bauthor{\bsnm{Peres}, \binits{A.}},
\bauthor{\bsnm{Wootters}, \binits{W.K.}}:
\batitle{{Teleporting an unknown quantum state via dual classical and Einstein-Podolsky-Rosen channels}}.
\bjtitle{Physical Review Letters}
\bvolume{70}(\bissue{13}),
\bfpage{1895}--\blpage{1899}
(\byear{1993})
\doiurl{10.1103/PhysRevLett.70.1895}
\end{barticle}
\endbibitem

\bibitem[\protect\citeauthoryear{{Grimau Puigibert} et~al.}{2017}]{GrimauPuigibert2017}
\begin{barticle}
\bauthor{\bsnm{{Grimau Puigibert}}, \binits{M.}},
\bauthor{\bsnm{Aguilar}, \binits{G.H.}},
\bauthor{\bsnm{Zhou}, \binits{Q.}},
\bauthor{\bsnm{Marsili}, \binits{F.}},
\bauthor{\bsnm{Shaw}, \binits{M.D.}},
\bauthor{\bsnm{Verma}, \binits{V.B.}},
\bauthor{\bsnm{Nam}, \binits{S.W.}},
\bauthor{\bsnm{Oblak}, \binits{D.}},
\bauthor{\bsnm{Tittel}, \binits{W.}}:
\batitle{{Heralded Single Photons Based on Spectral Multiplexing and Feed-Forward Control}}.
\bjtitle{Physical Review Letters}
\bvolume{119}(\bissue{8}),
\bfpage{083601}
(\byear{2017})
\doiurl{10.1103/PhysRevLett.119.083601}
\end{barticle}
\endbibitem

\bibitem[\protect\citeauthoryear{Kaneda and Kwiat}{2019}]{Kaneda2019}
\begin{barticle}
\bauthor{\bsnm{Kaneda}, \binits{F.}},
\bauthor{\bsnm{Kwiat}, \binits{P.G.}}:
\batitle{{High-efficiency single-photon generation via large-scale active time multiplexing}}.
\bjtitle{Science Advances}
\bvolume{5}(\bissue{10}),
\bfpage{1}--\blpage{6}
(\byear{2019})
\doiurl{10.1126/sciadv.aaw8586}
{\href{https://arxiv.org/abs/1803.04803}{{1803.04803}}}
\end{barticle}
\endbibitem

\bibitem[\protect\citeauthoryear{Hiemstra et~al.}{2020}]{Hiemstra2020}
\begin{barticle}
\bauthor{\bsnm{Hiemstra}, \binits{T.}},
\bauthor{\bsnm{Parker}, \binits{T.F.}},
\bauthor{\bsnm{Humphreys}, \binits{P.}},
\bauthor{\bsnm{Tiedau}, \binits{J.}},
\bauthor{\bsnm{Beck}, \binits{M.}},
\bauthor{\bsnm{Karpi{\'{n}}ski}, \binits{M.}},
\bauthor{\bsnm{Smith}, \binits{B.J.}},
\bauthor{\bsnm{Eckstein}, \binits{A.}},
\bauthor{\bsnm{Kolthammer}, \binits{W.S.}},
\bauthor{\bsnm{Walmsley}, \binits{I.A.}}:
\batitle{{Pure Single Photons from Scalable Frequency Multiplexing}}.
\bjtitle{Physical Review Applied}
\bvolume{14}(\bissue{1}),
\bfpage{1}--\blpage{10}
(\byear{2020})
\doiurl{10.1103/PhysRevApplied.14.014052}
{\href{https://arxiv.org/abs/1907.10355}{{1907.10355}}}
\end{barticle}
\endbibitem

\bibitem[\protect\citeauthoryear{Pittman et~al.}{2002}]{Pittman2002}
\begin{barticle}
\bauthor{\bsnm{Pittman}, \binits{T.B.}},
\bauthor{\bsnm{Jacobs}, \binits{B.C.}},
\bauthor{\bsnm{Franson}, \binits{J.D.}}:
\batitle{{Demonstration of feed-forward control for linear optics quantum computation}}.
\bjtitle{Physical Review A - Atomic, Molecular, and Optical Physics}
\bvolume{66}(\bissue{5}),
\bfpage{7}
(\byear{2002})
\doiurl{10.1103/PhysRevA.66.052305}
{\href{https://arxiv.org/abs/0204142}{{0204142}}}
\end{barticle}
\endbibitem

\bibitem[\protect\citeauthoryear{Prevedel et~al.}{2007}]{Prevedel2007}
\begin{barticle}
\bauthor{\bsnm{Prevedel}, \binits{R.}},
\bauthor{\bsnm{Walther}, \binits{P.}},
\bauthor{\bsnm{Tiefenbacher}, \binits{F.}},
\bauthor{\bsnm{B{\"{o}}hi}, \binits{P.}},
\bauthor{\bsnm{Kaltenbaek}, \binits{R.}},
\bauthor{\bsnm{Jennewein}, \binits{T.}},
\bauthor{\bsnm{Zeilinger}, \binits{A.}}:
\batitle{{High-speed linear optics quantum computing using active feed-forward}}.
\bjtitle{Nature}
\bvolume{445}(\bissue{7123}),
\bfpage{65}--\blpage{69}
(\byear{2007})
\doiurl{10.1038/nature05346}
\end{barticle}
\endbibitem

\bibitem[\protect\citeauthoryear{Giacomini et~al.}{2002}]{Giacomini2002}
\begin{barticle}
\bauthor{\bsnm{Giacomini}, \binits{S.}},
\bauthor{\bsnm{Sciarrino}, \binits{F.}},
\bauthor{\bsnm{Lombardi}, \binits{E.}},
\bauthor{\bsnm{{De Martini}}, \binits{F.}}:
\batitle{{Active teleportation of a quantum bit}}.
\bjtitle{Physical Review A - Atomic, Molecular, and Optical Physics}
\bvolume{66}(\bissue{3}),
\bfpage{4}
(\byear{2002})
\doiurl{10.1103/PhysRevA.66.030302}
\end{barticle}
\endbibitem

\bibitem[\protect\citeauthoryear{Pirandola et~al.}{2015}]{Pirandola2015}
\begin{barticle}
\bauthor{\bsnm{Pirandola}, \binits{S.}},
\bauthor{\bsnm{Eisert}, \binits{J.}},
\bauthor{\bsnm{Weedbrook}, \binits{C.}},
\bauthor{\bsnm{Furusawa}, \binits{A.}},
\bauthor{\bsnm{Braunstein}, \binits{S.L.}}:
\batitle{{Advances in quantum teleportation}}.
\bjtitle{Nature Photonics}
\bvolume{9}(\bissue{10}),
\bfpage{641}--\blpage{652}
(\byear{2015})
\doiurl{10.1038/nphoton.2015.154}
{\href{https://arxiv.org/abs/1505.07831}{{1505.07831}}}
\end{barticle}
\endbibitem

\bibitem[\protect\citeauthoryear{Pittman and Franson}{2002}]{Pittman2002a}
\begin{barticle}
\bauthor{\bsnm{Pittman}, \binits{T.B.}},
\bauthor{\bsnm{Franson}, \binits{J.D.}}:
\batitle{{Cyclical quantum memory for photonic qubits}}.
\bjtitle{Physical Review A}
\bvolume{66}(\bissue{6}),
\bfpage{062302}
(\byear{2002})
\doiurl{10.1103/PhysRevA.66.062302}
\end{barticle}
\endbibitem

\bibitem[\protect\citeauthoryear{Massaro et~al.}{2019}]{Massaro2019}
\begin{barticle}
\bauthor{\bsnm{Massaro}, \binits{M.}},
\bauthor{\bsnm{Meyer-Scott}, \binits{E.}},
\bauthor{\bsnm{Montaut}, \binits{N.}},
\bauthor{\bsnm{Herrmann}, \binits{H.}},
\bauthor{\bsnm{Silberhorn}, \binits{C.}}:
\batitle{{Improving SPDC single-photon sources via extended heralding and feed-forward control}}.
\bjtitle{New Journal of Physics}
\bvolume{21}(\bissue{5}),
\bfpage{053038}
(\byear{2019})
\doiurl{10.1088/1367-2630/ab1ec3}
\end{barticle}
\endbibitem

\bibitem[\protect\citeauthoryear{{Luiz Zanin} et~al.}{2021}]{LuizZanin2021}
\begin{barticle}
\bauthor{\bsnm{{Luiz Zanin}}, \binits{G.}},
\bauthor{\bsnm{Jacquet}, \binits{M.J.}},
\bauthor{\bsnm{Spagnolo}, \binits{M.}},
\bauthor{\bsnm{Schiansky}, \binits{P.}},
\bauthor{\bsnm{Calafell}, \binits{I.A.}},
\bauthor{\bsnm{Rozema}, \binits{L.A.}},
\bauthor{\bsnm{Walther}, \binits{P.}}:
\batitle{{Fiber-compatible photonic feed-forward with 99{\%} fidelity}}.
\bjtitle{Optics Express}
\bvolume{29}(\bissue{3}),
\bfpage{3425}
(\byear{2021})
\doiurl{10.1364/oe.409867}
{\href{https://arxiv.org/abs/2009.07868}{{2009.07868}}}
\end{barticle}
\endbibitem

\bibitem[\protect\citeauthoryear{Meyer-Scott et~al.}{2022}]{Meyer-scott2022}
\begin{barticle}
\bauthor{\bsnm{Meyer-Scott}, \binits{E.}},
\bauthor{\bsnm{Prasannan}, \binits{N.}},
\bauthor{\bsnm{Dhand}, \binits{I.}},
\bauthor{\bsnm{Eigner}, \binits{C.}},
\bauthor{\bsnm{Quiring}, \binits{V.}},
\bauthor{\bsnm{Barkhofen}, \binits{S.}},
\bauthor{\bsnm{Brecht}, \binits{B.}},
\bauthor{\bsnm{Plenio}, \binits{M.B.}},
\bauthor{\bsnm{Silberhorn}, \binits{C.}}:
\batitle{{Scalable Generation of Multiphoton Entangled States by Active Feed-Forward and Multiplexing}}.
\bjtitle{Physical Review Letters}
\bvolume{129}(\bissue{15}),
\bfpage{150501}
(\byear{2022})
\doiurl{10.1103/PhysRevLett.129.150501}
{\href{https://arxiv.org/abs/1908.05722}{{1908.05722}}}
\end{barticle}
\endbibitem

\bibitem[\protect\citeauthoryear{Hou et~al.}{2023}]{Hou2023}
\begin{barticle}
\bauthor{\bsnm{Hou}, \binits{Z.}},
\bauthor{\bsnm{Tang}, \binits{J.F.}},
\bauthor{\bsnm{Huang}, \binits{C.J.}},
\bauthor{\bsnm{Huang}, \binits{Y.F.}},
\bauthor{\bsnm{Xiang}, \binits{G.Y.}},
\bauthor{\bsnm{Li}, \binits{C.F.}},
\bauthor{\bsnm{Guo}, \binits{G.C.}}:
\batitle{{Entangled-State Time Multiplexing for Multiphoton Entanglement Generation}}.
\bjtitle{Physical Review Applied}
\bvolume{19}(\bissue{1}),
\bfpage{1}
(\byear{2023})
\doiurl{10.1103/PhysRevApplied.19.L011002}
\end{barticle}
\endbibitem

\bibitem[\protect\citeauthoryear{Donaldson et~al.}{2019}]{Donaldson2019}
\begin{barticle}
\bauthor{\bsnm{Donaldson}, \binits{R.J.}},
\bauthor{\bsnm{Mazzarella}, \binits{L.}},
\bauthor{\bsnm{Zanforlin}, \binits{U.}},
\bauthor{\bsnm{Collins}, \binits{R.J.}},
\bauthor{\bsnm{Jeffers}, \binits{J.}},
\bauthor{\bsnm{Buller}, \binits{G.S.}}:
\batitle{{Quantum state correction using a measurement-based feedforward mechanism}}.
\bjtitle{Physical Review A}
\bvolume{100}(\bissue{2}),
\bfpage{23840}
(\byear{2019})
\doiurl{10.1103/PhysRevA.100.023840}
\end{barticle}
\endbibitem

\bibitem[\protect\citeauthoryear{Sabines-Chesterking et~al.}{2017}]{Sabines-chesterking2017}
\begin{barticle}
\bauthor{\bsnm{Sabines-Chesterking}, \binits{J.}},
\bauthor{\bsnm{Whittaker}, \binits{R.}},
\bauthor{\bsnm{Joshi}, \binits{S.K.}},
\bauthor{\bsnm{Birchall}, \binits{P.M.}},
\bauthor{\bsnm{Moreau}, \binits{P.A.}},
\bauthor{\bsnm{McMillan}, \binits{A.}},
\bauthor{\bsnm{Cable}, \binits{H.V.}},
\bauthor{\bsnm{O'Brien}, \binits{J.L.}},
\bauthor{\bsnm{Rarity}, \binits{J.G.}},
\bauthor{\bsnm{Matthews}, \binits{J.C.F.}}:
\batitle{{Sub-Shot-Noise Transmission Measurement Enabled by Active Feed-Forward of Heralded Single Photons}}.
\bjtitle{Physical Review Applied}
\bvolume{8}(\bissue{1}),
\bfpage{6}--\blpage{11}
(\byear{2017})
\doiurl{10.1103/PhysRevApplied.8.014016}
\end{barticle}
\endbibitem

\bibitem[\protect\citeauthoryear{Mikov{\'{a}} et~al.}{2012}]{Mikova2012}
\begin{barticle}
\bauthor{\bsnm{Mikov{\'{a}}}, \binits{M.}},
\bauthor{\bsnm{Fikerov{\'{a}}}, \binits{H.}},
\bauthor{\bsnm{Straka}, \binits{I.}},
\bauthor{\bsnm{Mi{\v{c}}uda}, \binits{M.}},
\bauthor{\bsnm{Fiur{\'{a}}{\v{s}}ek}, \binits{J.}},
\bauthor{\bsnm{Je{\v{z}}ek}, \binits{M.}},
\bauthor{\bsnm{Du{\v{s}}ek}, \binits{M.}}:
\batitle{{Increasing efficiency of a linear-optical quantum gate using electronic feed-forward}}.
\bjtitle{Physical Review A - Atomic, Molecular, and Optical Physics}
\bvolume{85}(\bissue{1}),
\bfpage{1}--\blpage{4}
(\byear{2012})
\doiurl{10.1103/PhysRevA.85.012305}
{\href{https://arxiv.org/abs/1111.3237}{{1111.3237}}}
\end{barticle}
\endbibitem

\bibitem[\protect\citeauthoryear{Reddy et~al.}{2020}]{Reddy2020}
\begin{barticle}
\bauthor{\bsnm{Reddy}, \binits{D.V.}},
\bauthor{\bsnm{Nerem}, \binits{R.R.}},
\bauthor{\bsnm{Nam}, \binits{S.W.}},
\bauthor{\bsnm{Mirin}, \binits{R.P.}},
\bauthor{\bsnm{Verma}, \binits{V.B.}}:
\batitle{{Superconducting nanowire single-photon detectors with 98$\backslash${\%} system detection efficiency at 1550 nm}}.
\bjtitle{Optica}
\bvolume{7}(\bissue{12}),
\bfpage{1649}
(\byear{2020})
\doiurl{10.1364/optica.400751}
\end{barticle}
\endbibitem

\bibitem[\protect\citeauthoryear{Korzh et~al.}{2020}]{Korzh2020}
\begin{barticle}
\bauthor{\bsnm{Korzh}, \binits{B.}},
\bauthor{\bsnm{Zhao}, \binits{Q.Y.}},
\bauthor{\bsnm{Allmaras}, \binits{J.P.}},
\bauthor{\bsnm{Frasca}, \binits{S.}},
\bauthor{\bsnm{Autry}, \binits{T.M.}},
\bauthor{\bsnm{Bersin}, \binits{E.A.}},
\bauthor{\bsnm{Beyer}, \binits{A.D.}},
\bauthor{\bsnm{Briggs}, \binits{R.M.}},
\bauthor{\bsnm{Bumble}, \binits{B.}},
\bauthor{\bsnm{Colangelo}, \binits{M.}},
\bauthor{\bsnm{Crouch}, \binits{G.M.}},
\bauthor{\bsnm{Dane}, \binits{A.E.}},
\bauthor{\bsnm{Gerrits}, \binits{T.}},
\bauthor{\bsnm{Lita}, \binits{A.E.}},
\bauthor{\bsnm{Marsili}, \binits{F.}},
\bauthor{\bsnm{Moody}, \binits{G.}},
\bauthor{\bsnm{Pe{\~{n}}a}, \binits{C.}},
\bauthor{\bsnm{Ramirez}, \binits{E.}},
\bauthor{\bsnm{Rezac}, \binits{J.D.}},
\bauthor{\bsnm{Sinclair}, \binits{N.}},
\bauthor{\bsnm{Stevens}, \binits{M.J.}},
\bauthor{\bsnm{Velasco}, \binits{A.E.}},
\bauthor{\bsnm{Verma}, \binits{V.B.}},
\bauthor{\bsnm{Wollman}, \binits{E.E.}},
\bauthor{\bsnm{Xie}, \binits{S.}},
\bauthor{\bsnm{Zhu}, \binits{D.}},
\bauthor{\bsnm{Hale}, \binits{P.D.}},
\bauthor{\bsnm{Spiropulu}, \binits{M.}},
\bauthor{\bsnm{Silverman}, \binits{K.L.}},
\bauthor{\bsnm{Mirin}, \binits{R.P.}},
\bauthor{\bsnm{Nam}, \binits{S.W.}},
\bauthor{\bsnm{Kozorezov}, \binits{A.G.}},
\bauthor{\bsnm{Shaw}, \binits{M.D.}},
\bauthor{\bsnm{Berggren}, \binits{K.K.}}:
\batitle{{Demonstration of sub-3 ps temporal resolution with a superconducting nanowire single-photon detector}}.
\bjtitle{Nature Photonics}
\bvolume{14}(\bissue{4}),
\bfpage{250}--\blpage{255}
(\byear{2020})
\doiurl{10.1038/s41566-020-0589-x}
{\href{https://arxiv.org/abs/1804.06839}{{1804.06839}}}
\end{barticle}
\endbibitem

\bibitem[\protect\citeauthoryear{Chiles et~al.}{2022}]{Chiles2022}
\begin{barticle}
\bauthor{\bsnm{Chiles}, \binits{J.}},
\bauthor{\bsnm{Charaev}, \binits{I.}},
\bauthor{\bsnm{Lasenby}, \binits{R.}},
\bauthor{\bsnm{Baryakhtar}, \binits{M.}},
\bauthor{\bsnm{Huang}, \binits{J.}},
\bauthor{\bsnm{Roshko}, \binits{A.}},
\bauthor{\bsnm{Burton}, \binits{G.}},
\bauthor{\bsnm{Colangelo}, \binits{M.}},
\bauthor{\bsnm{{Van Tilburg}}, \binits{K.}},
\bauthor{\bsnm{Arvanitaki}, \binits{A.}},
\bauthor{\bsnm{Nam}, \binits{S.W.}},
\bauthor{\bsnm{Berggren}, \binits{K.K.}}:
\batitle{{New Constraints on Dark Photon Dark Matter with Superconducting Nanowire Detectors in an Optical Haloscope}}.
\bjtitle{Physical Review Letters}
\bvolume{128}(\bissue{23}),
\bfpage{231802}
(\byear{2022})
\doiurl{10.1103/PhysRevLett.128.231802}
{\href{https://arxiv.org/abs/2110.01582}{{2110.01582}}}
\end{barticle}
\endbibitem

\bibitem[\protect\citeauthoryear{Natarajan et~al.}{2012}]{Natarajan2012c}
\begin{botherref}
\oauthor{\bsnm{Natarajan}, \binits{C.M.}},
\oauthor{\bsnm{Tanner}, \binits{M.G.}},
\oauthor{\bsnm{Hadfield}, \binits{R.H.}}:
{Superconducting nanowire single-photon detectors: Physics and applications}.
Superconductor Science and Technology
\textbf{25}(6)
(2012)
\doiurl{10.1088/0953-2048/25/6/063001}
\end{botherref}
\endbibitem

\bibitem[\protect\citeauthoryear{de~Cea et~al.}{2020}]{DeCea2020}
\begin{barticle}
\bauthor{\bsnm{Cea}, \binits{M.}},
\bauthor{\bsnm{Wollman}, \binits{E.E.}},
\bauthor{\bsnm{Atabaki}, \binits{A.H.}},
\bauthor{\bsnm{Gray}, \binits{D.J.}},
\bauthor{\bsnm{Shaw}, \binits{M.D.}},
\bauthor{\bsnm{Ram}, \binits{R.J.}}:
\batitle{{Photonic Readout of Superconducting Nanowire Single Photon Counting Detectors}}.
\bjtitle{Scientific Reports}
\bvolume{10}(\bissue{1}),
\bfpage{9470}
(\byear{2020})
\doiurl{10.1038/s41598-020-65971-5}
\end{barticle}
\endbibitem

\bibitem[\protect\citeauthoryear{Thiele et~al.}{2023}]{Thiele2023}
\begin{barticle}
\bauthor{\bsnm{Thiele}, \binits{F.}},
\bauthor{\bsnm{Hummel}, \binits{T.}},
\bauthor{\bsnm{McCaughan}, \binits{A.N.}},
\bauthor{\bsnm{Brockmeier}, \binits{J.}},
\bauthor{\bsnm{Protte}, \binits{M.}},
\bauthor{\bsnm{Quiring}, \binits{V.}},
\bauthor{\bsnm{Lengeling}, \binits{S.}},
\bauthor{\bsnm{Eigner}, \binits{C.}},
\bauthor{\bsnm{Silberhorn}, \binits{C.}},
\bauthor{\bsnm{Bartley}, \binits{T.J.}}:
\batitle{{All optical operation of a superconducting photonic interface}}.
\bjtitle{Optics Express}
\bvolume{31}(\bissue{20}),
\bfpage{32717}
(\byear{2023})
\doiurl{10.1364/oe.492035}
{\href{https://arxiv.org/abs/2302.12123}{{2302.12123}}}
\end{barticle}
\endbibitem

\bibitem[\protect\citeauthoryear{Youssefi et~al.}{2021}]{Youssefi2021}
\begin{barticle}
\bauthor{\bsnm{Youssefi}, \binits{A.}},
\bauthor{\bsnm{Shomroni}, \binits{I.}},
\bauthor{\bsnm{Joshi}, \binits{Y.J.}},
\bauthor{\bsnm{Bernier}, \binits{N.R.}},
\bauthor{\bsnm{Lukashchuk}, \binits{A.}},
\bauthor{\bsnm{Uhrich}, \binits{P.}},
\bauthor{\bsnm{Qiu}, \binits{L.}},
\bauthor{\bsnm{Kippenberg}, \binits{T.J.}}:
\batitle{{A cryogenic electro-optic interconnect for superconducting devices}}.
\bjtitle{Nature Electronics}
\bvolume{4}(\bissue{5}),
\bfpage{326}--\blpage{332}
(\byear{2021})
\doiurl{10.1038/s41928-021-00570-4}
{\href{https://arxiv.org/abs/2004.04705}{{2004.04705}}}
\end{barticle}
\endbibitem

\bibitem[\protect\citeauthoryear{Thiele et~al.}{2022}]{Thiele2022c}
\begin{barticle}
\bauthor{\bsnm{Thiele}, \binits{F.}},
\bauthor{\bsnm{Hummel}, \binits{T.}},
\bauthor{\bsnm{Protte}, \binits{M.}},
\bauthor{\bsnm{Bartley}, \binits{T.J.}}:
\batitle{{Opto-electronic bias of a superconducting nanowire single photon detector using a cryogenic photodiode}}.
\bjtitle{APL Photonics}
\bvolume{7}(\bissue{8}),
\bfpage{081303}
(\byear{2022})
\doiurl{10.1063/5.0097506}
\end{barticle}
\endbibitem

\bibitem[\protect\citeauthoryear{Shen et~al.}{2024}]{Shen2024}
\begin{barticle}
\bauthor{\bsnm{Shen}, \binits{M.}},
\bauthor{\bsnm{Xie}, \binits{J.}},
\bauthor{\bsnm{Xu}, \binits{Y.}},
\bauthor{\bsnm{Wang}, \binits{S.}},
\bauthor{\bsnm{Cheng}, \binits{R.}},
\bauthor{\bsnm{Fu}, \binits{W.}},
\bauthor{\bsnm{Zhou}, \binits{Y.}},
\bauthor{\bsnm{Tang}, \binits{H.X.}}:
\batitle{{Photonic link from single-flux-quantum circuits to room temperature}}.
\bjtitle{Nature Photonics}
\bvolume{18}(\bissue{4}),
\bfpage{371}--\blpage{378}
(\byear{2024})
\doiurl{10.1038/s41566-023-01370-2}
{\href{https://arxiv.org/abs/2309.03284}{{2309.03284}}}
\end{barticle}
\endbibitem

\bibitem[\protect\citeauthoryear{Thiele et~al.}{2024}]{Thiele2024b}
\begin{botherref}
\oauthor{\bsnm{Thiele}, \binits{F.}},
\oauthor{\bsnm{Lamberty}, \binits{N.}},
\oauthor{\bsnm{Hummel}, \binits{T.}},
\oauthor{\bsnm{Bartley}, \binits{T.}}:
{Optical bias and cryogenic laser readout of a multipixel superconducting nanowire single photon detector}.
APL Photonics
\textbf{9}(7)
(2024)
\doiurl{10.1063/5.0209458}
{\href{https://arxiv.org/abs/2403.14276}{{arXiv:2403.14276}}}
\end{botherref}
\endbibitem

\bibitem[\protect\citeauthoryear{Sharapova et~al.}{2017}]{Sharapova2017b}
\begin{botherref}
\oauthor{\bsnm{Sharapova}, \binits{P.R.}},
\oauthor{\bsnm{Luo}, \binits{K.H.}},
\oauthor{\bsnm{Herrmann}, \binits{H.}},
\oauthor{\bsnm{Reichelt}, \binits{M.}},
\oauthor{\bsnm{Meier}, \binits{T.}},
\oauthor{\bsnm{Silberhorn}, \binits{C.}}:
{Toolbox for the design of LiNbO3-based passive and active integrated quantum circuits}.
New Journal of Physics
\textbf{19}(12)
(2017)
\doiurl{10.1088/1367-2630/aa9033}
{\href{https://arxiv.org/abs/1704.03769}{{arXiv:1704.03769}}}
\end{botherref}
\endbibitem

\bibitem[\protect\citeauthoryear{Luo et~al.}{2019}]{Luo2019}
\begin{barticle}
\bauthor{\bsnm{Luo}, \binits{K.-H.}},
\bauthor{\bsnm{Brauner}, \binits{S.}},
\bauthor{\bsnm{Eigner}, \binits{C.}},
\bauthor{\bsnm{Sharapova}, \binits{P.R.}},
\bauthor{\bsnm{Ricken}, \binits{R.}},
\bauthor{\bsnm{Meier}, \binits{T.}},
\bauthor{\bsnm{Herrmann}, \binits{H.}},
\bauthor{\bsnm{Silberhorn}, \binits{C.}}:
\batitle{{Nonlinear integrated quantum electro-optic circuits}}.
\bjtitle{Science Advances}
\bvolume{5}(\bissue{1}),
\bfpage{1}--\blpage{7}
(\byear{2019})
\doiurl{10.1126/sciadv.aat1451}
{\href{https://arxiv.org/abs/1810.13173}{{arXiv:1810.13173}}}
\end{barticle}
\endbibitem

\bibitem[\protect\citeauthoryear{Bartnick et~al.}{2021}]{Bartnick2021}
\begin{barticle}
\bauthor{\bsnm{Bartnick}, \binits{M.}},
\bauthor{\bsnm{Santandrea}, \binits{M.}},
\bauthor{\bsnm{H{\"{o}}pker}, \binits{J.P.}},
\bauthor{\bsnm{Thiele}, \binits{F.}},
\bauthor{\bsnm{Ricken}, \binits{R.}},
\bauthor{\bsnm{Quiring}, \binits{V.}},
\bauthor{\bsnm{Eigner}, \binits{C.}},
\bauthor{\bsnm{Herrmann}, \binits{H.}},
\bauthor{\bsnm{Silberhorn}, \binits{C.}},
\bauthor{\bsnm{Bartley}, \binits{T.J.}}:
\batitle{{Cryogenic Second-Harmonic Generation in Periodically Poled Lithium Niobate Waveguides}}.
\bjtitle{Physical Review Applied}
\bvolume{15}(\bissue{2}),
\bfpage{024028}
(\byear{2021})
\doiurl{10.1103/PhysRevApplied.15.024028}
\end{barticle}
\endbibitem

\bibitem[\protect\citeauthoryear{Lange et~al.}{2022}]{Lange2022}
\begin{barticle}
\bauthor{\bsnm{Lange}, \binits{N.A.}},
\bauthor{\bsnm{H{\"{o}}pker}, \binits{J.P.}},
\bauthor{\bsnm{Ricken}, \binits{R.}},
\bauthor{\bsnm{Quiring}, \binits{V.}},
\bauthor{\bsnm{Eigner}, \binits{C.}},
\bauthor{\bsnm{Silberhorn}, \binits{C.}},
\bauthor{\bsnm{Bartley}, \binits{T.J.}}:
\batitle{{Cryogenic integrated spontaneous parametric down-conversion}}.
\bjtitle{Optica}
\bvolume{9}(\bissue{1}),
\bfpage{108}
(\byear{2022})
\doiurl{10.1364/optica.445576}
{\href{https://arxiv.org/abs/2110.07425}{{arXiv:2110.07425}}}
\end{barticle}
\endbibitem

\bibitem[\protect\citeauthoryear{Lange et~al.}{2023}]{Lange2023}
\begin{barticle}
\bauthor{\bsnm{Lange}, \binits{N.A.}},
\bauthor{\bsnm{Schapeler}, \binits{T.}},
\bauthor{\bsnm{H{\"{o}}pker}, \binits{J.P.}},
\bauthor{\bsnm{Protte}, \binits{M.}},
\bauthor{\bsnm{Bartley}, \binits{T.J.}}:
\batitle{{Degenerate photons from a cryogenic spontaneous parametric down-conversion source}}.
\bjtitle{Physical Review A}
\bvolume{108}(\bissue{2}),
\bfpage{023701}
(\byear{2023})
\doiurl{10.1103/PhysRevA.108.023701}
{\href{https://arxiv.org/abs/2303.17428}{{arXiv:2303.17428}}}
\end{barticle}
\endbibitem

\bibitem[\protect\citeauthoryear{Yoshida et~al.}{1999}]{Yoshida1999a}
\begin{barticle}
\bauthor{\bsnm{Yoshida}, \binits{K.}},
\bauthor{\bsnm{Kanda}, \binits{Y.}},
\bauthor{\bsnm{Kohjiro}, \binits{S.}}:
\batitle{{A traveling-wave-type LiNbO3 optical modulator with superconducting electrodes}}.
\bjtitle{IEEE Transactions on Microwave Theory and Techniques}
\bvolume{47}(\bissue{7}),
\bfpage{1201}--\blpage{1205}
(\byear{1999})
\doiurl{10.1109/22.775458}
\end{barticle}
\endbibitem

\bibitem[\protect\citeauthoryear{Thiele et~al.}{2020}]{Thiele2020}
\begin{barticle}
\bauthor{\bsnm{Thiele}, \binits{F.}},
\bauthor{\bsnm{Bruch}, \binits{F.}},
\bauthor{\bsnm{Quiring}, \binits{V.}},
\bauthor{\bsnm{Ricken}, \binits{R.}},
\bauthor{\bsnm{Herrmann}, \binits{H.}},
\bauthor{\bsnm{Eigner}, \binits{C.}},
\bauthor{\bsnm{Silberhorn}, \binits{C.}},
\bauthor{\bsnm{Bartley}, \binits{T.J.}}:
\batitle{{Cryogenic electro-optic polarisation conversion in titanium in-diffused lithium niobate waveguides}}.
\bjtitle{Optics Express}
\bvolume{28}(\bissue{20}),
\bfpage{28961}
(\byear{2020})
\doiurl{10.1364/OE.399818}
\end{barticle}
\endbibitem

\bibitem[\protect\citeauthoryear{Thiele et~al.}{2022}]{Thiele2022}
\begin{barticle}
\bauthor{\bsnm{Thiele}, \binits{F.}},
\bauthor{\bsnm{{Vom Bruch}}, \binits{F.}},
\bauthor{\bsnm{Brockmeier}, \binits{J.}},
\bauthor{\bsnm{Protte}, \binits{M.}},
\bauthor{\bsnm{Hummel}, \binits{T.}},
\bauthor{\bsnm{Ricken}, \binits{R.}},
\bauthor{\bsnm{Quiring}, \binits{V.}},
\bauthor{\bsnm{Lengeling}, \binits{S.}},
\bauthor{\bsnm{Herrmann}, \binits{H.}},
\bauthor{\bsnm{Eigner}, \binits{C.}},
\bauthor{\bsnm{Silberhorn}, \binits{C.}},
\bauthor{\bsnm{Bartley}, \binits{T.J.}}:
\batitle{{Cryogenic electro-optic modulation in titanium in-diffused lithium niobate waveguides}}.
\bjtitle{JPhys Photonics}
\bvolume{4}(\bissue{3}),
\bfpage{28961}--\blpage{28968}
(\byear{2022})
\doiurl{10.1088/2515-7647/ac6c63}
{\href{https://arxiv.org/abs/2202.00306}{{2202.00306}}}
\end{barticle}
\endbibitem

\bibitem[\protect\citeauthoryear{Lomonte et~al.}{2021}]{Lomonte2021b}
\begin{barticle}
\bauthor{\bsnm{Lomonte}, \binits{E.}},
\bauthor{\bsnm{Wolff}, \binits{M.A.}},
\bauthor{\bsnm{Beutel}, \binits{F.}},
\bauthor{\bsnm{Ferrari}, \binits{S.}},
\bauthor{\bsnm{Schuck}, \binits{C.}},
\bauthor{\bsnm{Pernice}, \binits{W.H.P.}},
\bauthor{\bsnm{Lenzini}, \binits{F.}}:
\batitle{{Single-photon detection and cryogenic reconfigurability in lithium niobate nanophotonic circuits}}.
\bjtitle{Nature Communications}
\bvolume{12}(\bissue{1}),
\bfpage{6847}
(\byear{2021})
\doiurl{10.1038/s41467-021-27205-8}
{\href{https://arxiv.org/abs/2103.10973}{{arXiv:2103.10973}}}
\end{barticle}
\endbibitem

\bibitem[\protect\citeauthoryear{Tanner et~al.}{2012}]{Tanner2012}
\begin{barticle}
\bauthor{\bsnm{Tanner}, \binits{M.G.}},
\bauthor{\bsnm{Alvarez}, \binits{L.S.E.}},
\bauthor{\bsnm{Jiang}, \binits{W.}},
\bauthor{\bsnm{Warburton}, \binits{R.J.}},
\bauthor{\bsnm{Barber}, \binits{Z.H.}},
\bauthor{\bsnm{Hadfield}, \binits{R.H.}}:
\batitle{{A superconducting nanowire single photon detector on lithium niobate}}.
\bjtitle{Nanotechnology}
\bvolume{23}(\bissue{50}),
\bfpage{505201}
(\byear{2012})
\doiurl{10.1088/0957-4484/23/50/505201}
\end{barticle}
\endbibitem

\bibitem[\protect\citeauthoryear{H{\"{o}}pker et~al.}{2019}]{Hoepker2019}
\begin{botherref}
\oauthor{\bsnm{H{\"{o}}pker}, \binits{J.P.}},
\oauthor{\bsnm{Gerrits}, \binits{T.}},
\oauthor{\bsnm{Lita}, \binits{A.}},
\oauthor{\bsnm{Krapick}, \binits{S.}},
\oauthor{\bsnm{Herrmann}, \binits{H.}},
\oauthor{\bsnm{Ricken}, \binits{R.}},
\oauthor{\bsnm{Quiring}, \binits{V.}},
\oauthor{\bsnm{Mirin}, \binits{R.}},
\oauthor{\bsnm{Nam}, \binits{S.W.}},
\oauthor{\bsnm{Silberhorn}, \binits{C.}},
\oauthor{\bsnm{Bartley}, \binits{T.J.}}:
{Integrated transition edge sensors on titanium in-diffused lithium niobate waveguides}.
APL Photonics
\textbf{4}(5)
(2019)
\doiurl{10.1063/1.5086276}
\end{botherref}
\endbibitem

\bibitem[\protect\citeauthoryear{H{\"{o}}pker et~al.}{2021}]{Hoepker2021b}
\begin{botherref}
\oauthor{\bsnm{H{\"{o}}pker}, \binits{J.P.}},
\oauthor{\bsnm{Verma}, \binits{V.B.}},
\oauthor{\bsnm{Protte}, \binits{M.}},
\oauthor{\bsnm{Ricken}, \binits{R.}},
\oauthor{\bsnm{Quiring}, \binits{V.}},
\oauthor{\bsnm{Eigner}, \binits{C.}},
\oauthor{\bsnm{Ebers}, \binits{L.}},
\oauthor{\bsnm{Hammer}, \binits{M.}},
\oauthor{\bsnm{F{\"{o}}rstner}, \binits{J.}},
\oauthor{\bsnm{Silberhorn}, \binits{C.}},
\oauthor{\bsnm{Mirin}, \binits{R.P.}},
\oauthor{\bsnm{Nam}, \binits{S.W.}},
\oauthor{\bsnm{Bartley}, \binits{T.J.}}:
{Integrated superconducting nanowire single-photon detectors on titanium in-diffused lithium niobate waveguides}.
JPhys Photonics
\textbf{3}(3)
(2021)
\doiurl{10.1088/2515-7647/ac105b}
{\href{https://arxiv.org/abs/2104.12500}{{arXiv:2104.12500}}}
\end{botherref}
\endbibitem

\bibitem[\protect\citeauthoryear{Gehl et~al.}{2017}]{Gehl2016a}
\begin{barticle}
\bauthor{\bsnm{Gehl}, \binits{M.}},
\bauthor{\bsnm{Long}, \binits{C.}},
\bauthor{\bsnm{Trotter}, \binits{D.}},
\bauthor{\bsnm{Starbuck}, \binits{A.}},
\bauthor{\bsnm{Pomerene}, \binits{A.}},
\bauthor{\bsnm{Wright}, \binits{J.B.}},
\bauthor{\bsnm{Melgaard}, \binits{S.}},
\bauthor{\bsnm{Siirola}, \binits{J.}},
\bauthor{\bsnm{Lentine}, \binits{A.L.}},
\bauthor{\bsnm{DeRose}, \binits{C.}}:
\batitle{{Operation of high-speed silicon photonic micro-disk modulators at cryogenic temperatures}}.
\bjtitle{Optica}
\bvolume{4}(\bissue{3}),
\bfpage{374}
(\byear{2017})
\doiurl{10.1364/OPTICA.4.000374}
\end{barticle}
\endbibitem

\bibitem[\protect\citeauthoryear{Gyger et~al.}{2021}]{Gyger2021}
\begin{barticle}
\bauthor{\bsnm{Gyger}, \binits{S.}},
\bauthor{\bsnm{Zichi}, \binits{J.}},
\bauthor{\bsnm{Schweickert}, \binits{L.}},
\bauthor{\bsnm{Elshaari}, \binits{A.W.}},
\bauthor{\bsnm{Steinhauer}, \binits{S.}},
\bauthor{\bsnm{{Covre da Silva}}, \binits{S.F.}},
\bauthor{\bsnm{Rastelli}, \binits{A.}},
\bauthor{\bsnm{Zwiller}, \binits{V.}},
\bauthor{\bsnm{J{\"{o}}ns}, \binits{K.D.}},
\bauthor{\bsnm{Errando-Herranz}, \binits{C.}}:
\batitle{{Reconfigurable photonics with on-chip single-photon detectors}}.
\bjtitle{Nature Communications}
\bvolume{12}(\bissue{1}),
\bfpage{1}--\blpage{8}
(\byear{2021})
\doiurl{10.1038/s41467-021-21624-3}
\end{barticle}
\endbibitem

\bibitem[\protect\citeauthoryear{Beutel et~al.}{2022}]{Beutel2022}
\begin{barticle}
\bauthor{\bsnm{Beutel}, \binits{F.}},
\bauthor{\bsnm{Grottke}, \binits{T.}},
\bauthor{\bsnm{Wolff}, \binits{M.A.}},
\bauthor{\bsnm{Schuck}, \binits{C.}},
\bauthor{\bsnm{Pernice}, \binits{W.H.P.}}:
\batitle{{Cryo-compatible opto-mechanical low-voltage phase-modulator integrated with superconducting single-photon detectors}}.
\bjtitle{Optics Express}
\bvolume{30}(\bissue{17}),
\bfpage{30066}
(\byear{2022})
\doiurl{10.1364/oe.462163}
\end{barticle}
\endbibitem

\bibitem[\protect\citeauthoryear{Alexander et~al.}{2024}]{Alexander2024}
\begin{botherref}
\oauthor{\bsnm{Alexander}, \binits{K.}},
\oauthor{\bsnm{Bahgat}, \binits{A.}},
\oauthor{\bsnm{Benyamini}, \binits{A.}},
\oauthor{\bsnm{Black}, \binits{D.}},
\oauthor{\bsnm{Bonneau}, \binits{D.}},
\oauthor{\bsnm{Burgos}, \binits{S.}},
\oauthor{\bsnm{Burridge}, \binits{B.}},
\oauthor{\bsnm{Campbell}, \binits{G.}},
\oauthor{\bsnm{Catalano}, \binits{G.}},
\oauthor{\bsnm{Ceballos}, \binits{A.}},
\oauthor{\bsnm{Chang}, \binits{C.-M.}},
\oauthor{\bsnm{Chung}, \binits{C.}},
\oauthor{\bsnm{Danesh}, \binits{F.}},
\oauthor{\bsnm{Dauer}, \binits{T.}},
\oauthor{\bsnm{Davis}, \binits{M.}},
\oauthor{\bsnm{Dudley}, \binits{E.}},
\oauthor{\bsnm{Er-Xuan}, \binits{P.}},
\oauthor{\bsnm{Fargas}, \binits{J.}},
\oauthor{\bsnm{Farsi}, \binits{A.}},
\oauthor{\bsnm{Fenrich}, \binits{C.}},
\oauthor{\bsnm{Frazer}, \binits{J.}},
\oauthor{\bsnm{Fukami}, \binits{M.}},
\oauthor{\bsnm{Ganesan}, \binits{Y.}},
\oauthor{\bsnm{Gibson}, \binits{G.}},
\oauthor{\bsnm{Gimeno-Segovia}, \binits{M.}},
\oauthor{\bsnm{Goeldi}, \binits{S.}},
\oauthor{\bsnm{Goley}, \binits{P.}},
\oauthor{\bsnm{Haislmaier}, \binits{R.}},
\oauthor{\bsnm{Halimi}, \binits{S.}},
\oauthor{\bsnm{Hansen}, \binits{P.}},
\oauthor{\bsnm{Hardy}, \binits{S.}},
\oauthor{\bsnm{Horng}, \binits{J.}},
\oauthor{\bsnm{House}, \binits{M.}},
\oauthor{\bsnm{Hu}, \binits{H.}},
\oauthor{\bsnm{Jadidi}, \binits{M.}},
\oauthor{\bsnm{Johansson}, \binits{H.}},
\oauthor{\bsnm{Jones}, \binits{T.}},
\oauthor{\bsnm{Kamineni}, \binits{V.}},
\oauthor{\bsnm{Kelez}, \binits{N.}},
\oauthor{\bsnm{Koustuban}, \binits{R.}},
\oauthor{\bsnm{Kovall}, \binits{G.}},
\oauthor{\bsnm{Krogen}, \binits{P.}},
\oauthor{\bsnm{Kumar}, \binits{N.}},
\oauthor{\bsnm{Liang}, \binits{Y.}},
\oauthor{\bsnm{LiCausi}, \binits{N.}},
\oauthor{\bsnm{Llewellyn}, \binits{D.}},
\oauthor{\bsnm{Lokovic}, \binits{K.}},
\oauthor{\bsnm{Lovelady}, \binits{M.}},
\oauthor{\bsnm{Manfrinato}, \binits{V.}},
\oauthor{\bsnm{Melnichuk}, \binits{A.}},
\oauthor{\bsnm{Souza}, \binits{M.}},
\oauthor{\bsnm{Mendoza}, \binits{G.}},
\oauthor{\bsnm{Moores}, \binits{B.}},
\oauthor{\bsnm{Mukherjee}, \binits{S.}},
\oauthor{\bsnm{Munns}, \binits{J.}},
\oauthor{\bsnm{Musalem}, \binits{F.-X.}},
\oauthor{\bsnm{Najafi}, \binits{F.}},
\oauthor{\bsnm{O'Brien}, \binits{J.L.}},
\oauthor{\bsnm{Ortmann}, \binits{J.E.}},
\oauthor{\bsnm{Pai}, \binits{S.}},
\oauthor{\bsnm{Park}, \binits{B.}},
\oauthor{\bsnm{Peng}, \binits{H.-T.}},
\oauthor{\bsnm{Penthorn}, \binits{N.}},
\oauthor{\bsnm{Peterson}, \binits{B.}},
\oauthor{\bsnm{Poush}, \binits{M.}},
\oauthor{\bsnm{Pryde}, \binits{G.J.}},
\oauthor{\bsnm{Ramprasad}, \binits{T.}},
\oauthor{\bsnm{Ray}, \binits{G.}},
\oauthor{\bsnm{Rodriguez}, \binits{A.}},
\oauthor{\bsnm{Roxworthy}, \binits{B.}},
\oauthor{\bsnm{Rudolph}, \binits{T.}},
\oauthor{\bsnm{Saunders}, \binits{D.J.}},
\oauthor{\bsnm{Shadbolt}, \binits{P.}},
\oauthor{\bsnm{Shah}, \binits{D.}},
\oauthor{\bsnm{Shin}, \binits{H.}},
\oauthor{\bsnm{Smith}, \binits{J.}},
\oauthor{\bsnm{Sohn}, \binits{B.}},
\oauthor{\bsnm{Sohn}, \binits{Y.-I.}},
\oauthor{\bsnm{Son}, \binits{G.}},
\oauthor{\bsnm{Sparrow}, \binits{C.}},
\oauthor{\bsnm{Staffaroni}, \binits{M.}},
\oauthor{\bsnm{Stavrakas}, \binits{C.}},
\oauthor{\bsnm{Sukumaran}, \binits{V.}},
\oauthor{\bsnm{Tamborini}, \binits{D.}},
\oauthor{\bsnm{Thompson}, \binits{M.G.}},
\oauthor{\bsnm{Tran}, \binits{K.}},
\oauthor{\bsnm{Triplet}, \binits{M.}},
\oauthor{\bsnm{Tung}, \binits{M.}},
\oauthor{\bsnm{Vert}, \binits{A.}},
\oauthor{\bsnm{Vidrighin}, \binits{M.D.}},
\oauthor{\bsnm{Vorobeichik}, \binits{I.}},
\oauthor{\bsnm{Weigel}, \binits{P.}},
\oauthor{\bsnm{Wingert}, \binits{M.}},
\oauthor{\bsnm{Wooding}, \binits{J.}},
\oauthor{\bsnm{Zhou}, \binits{X.}}:
{A manufacturable platform for photonic quantum computing}
(2024)
{\href{https://arxiv.org/abs/2404.17570}{{arXiv:2404.17570}}}
\end{botherref}
\endbibitem

\bibitem[\protect\citeauthoryear{Wolff et~al.}{2021}]{Wolff2021}
\begin{botherref}
\oauthor{\bsnm{Wolff}, \binits{M.A.}},
\oauthor{\bsnm{Beutel}, \binits{F.}},
\oauthor{\bsnm{Sch{\"{u}}tte}, \binits{J.}},
\oauthor{\bsnm{Gehring}, \binits{H.}},
\oauthor{\bsnm{H{\"{a}}u{\ss}ler}, \binits{M.}},
\oauthor{\bsnm{Pernice}, \binits{W.}},
\oauthor{\bsnm{Schuck}, \binits{C.}}:
{Broadband waveguide-integrated superconducting single-photon detectors with high system detection efficiency}.
Applied Physics Letters
\textbf{118}(15)
(2021)
\doiurl{10.1063/5.0046057}
\end{botherref}
\endbibitem

\bibitem[\protect\citeauthoryear{Wasserman et~al.}{2022}]{Wasserman2022}
\begin{barticle}
\bauthor{\bsnm{Wasserman}, \binits{W.W.}},
\bauthor{\bsnm{Harrison}, \binits{R.A.}},
\bauthor{\bsnm{Harris}, \binits{G.I.}},
\bauthor{\bsnm{Sawadsky}, \binits{A.}},
\bauthor{\bsnm{Sfendla}, \binits{Y.L.}},
\bauthor{\bsnm{Bowen}, \binits{W.P.}},
\bauthor{\bsnm{Baker}, \binits{C.G.}}:
\batitle{{Cryogenic and hermetically sealed packaging of photonic chips for optomechanics}}.
\bjtitle{Optics Express}
\bvolume{30}(\bissue{17}),
\bfpage{30822}
(\byear{2022})
\doiurl{10.1364/OE.463752}
{\href{https://arxiv.org/abs/2205.14143}{{arXiv:2205.14143}}}
\end{barticle}
\endbibitem

\bibitem[\protect\citeauthoryear{Lin et~al.}{2023}]{Lin2023}
\begin{botherref}
\oauthor{\bsnm{Lin}, \binits{B.}},
\oauthor{\bsnm{Witt}, \binits{D.}},
\oauthor{\bsnm{Young}, \binits{J.F.}},
\oauthor{\bsnm{Chrostowski}, \binits{L.}}:
{Cryogenic optical packaging using photonic wire bonds}.
APL Photonics
\textbf{8}(12)
(2023)
\doiurl{10.1063/5.0170974}
{\href{https://arxiv.org/abs/2307.07496}{{arXiv:2307.07496}}}
\end{botherref}
\endbibitem

\bibitem[\protect\citeauthoryear{Zeng et~al.}{2023}]{Zeng2023}
\begin{botherref}
\oauthor{\bsnm{Zeng}, \binits{B.}},
\oauthor{\bsnm{De-Eknamkul}, \binits{C.}},
\oauthor{\bsnm{Assumpcao}, \binits{D.}},
\oauthor{\bsnm{Renaud}, \binits{D.}},
\oauthor{\bsnm{Wang}, \binits{Z.}},
\oauthor{\bsnm{Riedel}, \binits{D.}},
\oauthor{\bsnm{Ha}, \binits{J.}},
\oauthor{\bsnm{Robens}, \binits{C.}},
\oauthor{\bsnm{Levonian}, \binits{D.}},
\oauthor{\bsnm{Lukin}, \binits{M.}},
\oauthor{\bsnm{Riedinger}, \binits{R.}},
\oauthor{\bsnm{Bhaskar}, \binits{M.}},
\oauthor{\bsnm{Sukachev}, \binits{D.}},
\oauthor{\bsnm{Loncar}, \binits{M.}},
\oauthor{\bsnm{Machielse}, \binits{B.}}:
{Cryogenic packaging of nanophotonic devices with a low coupling loss {\textless}1 dB}.
Applied Physics Letters
\textbf{123}(16)
(2023)
\doiurl{10.1063/5.0170324}
{\href{https://arxiv.org/abs/2306.09894}{{arXiv:2306.09894}}}
\end{botherref}
\endbibitem

\bibitem[\protect\citeauthoryear{Tiedau et~al.}{2020}]{Tiedau2020}
\begin{barticle}
\bauthor{\bsnm{Tiedau}, \binits{J.}},
\bauthor{\bsnm{Schapeler}, \binits{T.}},
\bauthor{\bsnm{Anant}, \binits{V.}},
\bauthor{\bsnm{Fedder}, \binits{H.}},
\bauthor{\bsnm{Silberhorn}, \binits{C.}},
\bauthor{\bsnm{Bartley}, \binits{T.J.}}:
\batitle{{Single-channel electronic readout of a multipixel superconducting nanowire single photon detector}}.
\bjtitle{Optics Express}
\bvolume{28}(\bissue{4}),
\bfpage{5528}
(\byear{2020})
\doiurl{10.1364/oe.383111}
{\href{https://arxiv.org/abs/1911.04880}{{1911.04880}}}
\end{barticle}
\endbibitem

\bibitem[\protect\citeauthoryear{Schapeler et~al.}{2020}]{Schapeler2020}
\begin{barticle}
\bauthor{\bsnm{Schapeler}, \binits{T.}},
\bauthor{\bsnm{{Philipp H{\"{o}}pker}}, \binits{J.}},
\bauthor{\bsnm{Bartley}, \binits{T.J.}}:
\batitle{{Quantum detector tomography of a 2×2 multi-pixel array of superconducting nanowire single photon detectors}}.
\bjtitle{Optics Express}
\bvolume{28}(\bissue{22}),
\bfpage{33035}
(\byear{2020})
\doiurl{10.1364/oe.404285}
{\href{https://arxiv.org/abs/2007.16048}{{2007.16048}}}
\end{barticle}
\endbibitem

\bibitem[\protect\citeauthoryear{Bashkansky et~al.}{2014}]{Bashkansky2014}
\begin{barticle}
\bauthor{\bsnm{Bashkansky}, \binits{M.}},
\bauthor{\bsnm{Vurgaftman}, \binits{I.}},
\bauthor{\bsnm{Pipino}, \binits{A.C.R.}},
\bauthor{\bsnm{Reintjes}, \binits{J.}}:
\batitle{{Significance of heralding in spontaneous parametric down-conversion}}.
\bjtitle{Physical Review A - Atomic, Molecular, and Optical Physics}
\bvolume{90}(\bissue{5}),
\bfpage{1}--\blpage{6}
(\byear{2014})
\doiurl{10.1103/PhysRevA.90.053825}
\end{barticle}
\endbibitem

\bibitem[\protect\citeauthoryear{Avenhaus et~al.}{2008}]{Avenhaus2008}
\begin{barticle}
\bauthor{\bsnm{Avenhaus}, \binits{M.}},
\bauthor{\bsnm{Coldenstrodt-Ronge}, \binits{H.B.}},
\bauthor{\bsnm{Laiho}, \binits{K.}},
\bauthor{\bsnm{Mauerer}, \binits{W.}},
\bauthor{\bsnm{Walmsley}, \binits{I.A.}},
\bauthor{\bsnm{Silberhorn}, \binits{C.}}:
\batitle{{Photon number statistics of multimode parametric down-conversion}}.
\bjtitle{Physical Review Letters}
\bvolume{101}(\bissue{5}),
\bfpage{1}--\blpage{4}
(\byear{2008})
\doiurl{10.1103/PhysRevLett.101.053601}
{\href{https://arxiv.org/abs/0804.0740}{{0804.0740}}}
\end{barticle}
\endbibitem

\bibitem[\protect\citeauthoryear{Stasi et~al.}{2024}]{Stasi2024}
\begin{botherref}
\oauthor{\bsnm{Stasi}, \binits{L.}},
\oauthor{\bsnm{Taher}, \binits{T.}},
\oauthor{\bsnm{Resta}, \binits{G.V.}},
\oauthor{\bsnm{Zbinden}, \binits{H.}},
\oauthor{\bsnm{Thew}, \binits{R.}},
\oauthor{\bsnm{Bussi{\`{e}}res}, \binits{F.}}:
{High photon-number efficiencies with a fast 28-pixel parallel SNSPD}
(2024)
{\href{https://arxiv.org/abs/2406.15312}{{2406.15312}}}
\end{botherref}
\endbibitem

\bibitem[\protect\citeauthoryear{Cahall et~al.}{2017}]{Cahall2017}
\begin{barticle}
\bauthor{\bsnm{Cahall}, \binits{C.}},
\bauthor{\bsnm{Nicolich}, \binits{K.L.}},
\bauthor{\bsnm{Islam}, \binits{N.T.}},
\bauthor{\bsnm{Lafyatis}, \binits{G.P.}},
\bauthor{\bsnm{Miller}, \binits{A.J.}},
\bauthor{\bsnm{Gauthier}, \binits{D.J.}},
\bauthor{\bsnm{Kim}, \binits{J.}}:
\batitle{{Multi-photon detection using a conventional superconducting nanowire single-photon detector}}.
\bjtitle{Optica}
\bvolume{4}(\bissue{12}),
\bfpage{1534}
(\byear{2017})
\doiurl{10.1364/optica.4.001534}
\end{barticle}
\endbibitem

\bibitem[\protect\citeauthoryear{Sauer et~al.}{2023}]{Sauer2023}
\begin{botherref}
\oauthor{\bsnm{Sauer}, \binits{G.}},
\oauthor{\bsnm{Kolarczik}, \binits{M.}},
\oauthor{\bsnm{Gomez}, \binits{R.}},
\oauthor{\bsnm{Conrad}, \binits{J.}},
\oauthor{\bsnm{Steinlechner}, \binits{F.}}:
{Resolving Photon Numbers Using Ultra-High-Resolution Timing of a Single Low-Jitter Superconducting Nanowire Detector}
(2023)
{\href{https://arxiv.org/abs/2310.12472}{{arXiv:2310.12472}}}
\end{botherref}
\endbibitem

\bibitem[\protect\citeauthoryear{Schapeler et~al.}{2024}]{Schapeler2024}
\begin{barticle}
\bauthor{\bsnm{Schapeler}, \binits{T.}},
\bauthor{\bsnm{Lamberty}, \binits{N.}},
\bauthor{\bsnm{Hummel}, \binits{T.}},
\bauthor{\bsnm{Schlue}, \binits{F.}},
\bauthor{\bsnm{Stefszky}, \binits{M.}},
\bauthor{\bsnm{Brecht}, \binits{B.}},
\bauthor{\bsnm{Silberhorn}, \binits{C.}},
\bauthor{\bsnm{Bartley}, \binits{T.J.}}:
\batitle{{Electrical trace analysis of superconducting nanowire photon-number-resolving detectors}}.
\bjtitle{Physical Review Applied}
\bvolume{22}(\bissue{1}),
\bfpage{1}
(\byear{2024})
\doiurl{10.1103/PhysRevApplied.22.014024}
\end{barticle}
\endbibitem

\bibitem[\protect\citeauthoryear{Kong et~al.}{2024}]{Kong2024}
\begin{barticle}
\bauthor{\bsnm{Kong}, \binits{L.-D.}},
\bauthor{\bsnm{Zhang}, \binits{T.-Z.}},
\bauthor{\bsnm{Liu}, \binits{X.-Y.}},
\bauthor{\bsnm{Li}, \binits{H.}},
\bauthor{\bsnm{Wang}, \binits{Z.}},
\bauthor{\bsnm{Xie}, \binits{X.-M.}},
\bauthor{\bsnm{You}, \binits{L.-X.}}:
\batitle{{Large-inductance superconducting microstrip photon detector enabling 10 photon-number resolution}}.
\bjtitle{Advanced Photonics}
\bvolume{6}(\bissue{01}),
\bfpage{1}--\blpage{10}
(\byear{2024})
\doiurl{10.1117/1.ap.6.1.016004}
\end{barticle}
\endbibitem

\bibitem[\protect\citeauthoryear{Krinner et~al.}{2019}]{Krinner2019}
\begin{botherref}
\oauthor{\bsnm{Krinner}, \binits{S.}},
\oauthor{\bsnm{Storz}, \binits{S.}},
\oauthor{\bsnm{Kurpiers}, \binits{P.}},
\oauthor{\bsnm{Magnard}, \binits{P.}},
\oauthor{\bsnm{Heinsoo}, \binits{J.}},
\oauthor{\bsnm{Keller}, \binits{R.}},
\oauthor{\bsnm{L{\"{u}}tolf}, \binits{J.}},
\oauthor{\bsnm{Eichler}, \binits{C.}},
\oauthor{\bsnm{Wallraff}, \binits{A.}}:
{Engineering cryogenic setups for 100-qubit scale superconducting circuit systems}.
EPJ Quantum Technology
\textbf{6}(1)
(2019)
\doiurl{10.1140/epjqt/s40507-019-0072-0}
{\href{https://arxiv.org/abs/1806.07862}{{arXiv:1806.07862}}}
\end{botherref}
\endbibitem

\bibitem[\protect\citeauthoryear{Lee et~al.}{2012}]{Lee2012}
\begin{botherref}
\oauthor{\bsnm{Lee}, \binits{H.}},
\oauthor{\bsnm{Chen}, \binits{T.}},
\oauthor{\bsnm{Li}, \binits{J.}},
\oauthor{\bsnm{Painter}, \binits{O.}},
\oauthor{\bsnm{Vahala}, \binits{K.J.}}:
{Ultra-low-loss optical delay line on a silicon chip}.
Nature Communications
\textbf{3}(May)
(2012)
\doiurl{10.1038/ncomms1876}
\end{botherref}
\endbibitem

\bibitem[\protect\citeauthoryear{Zhu et~al.}{2021}]{Zhu2021}
\begin{barticle}
\bauthor{\bsnm{Zhu}, \binits{D.}},
\bauthor{\bsnm{Shao}, \binits{L.}},
\bauthor{\bsnm{Yu}, \binits{M.}},
\bauthor{\bsnm{Cheng}, \binits{R.}},
\bauthor{\bsnm{Desiatov}, \binits{B.}},
\bauthor{\bsnm{Xin}, \binits{C.J.}},
\bauthor{\bsnm{Hu}, \binits{Y.}},
\bauthor{\bsnm{Holzgrafe}, \binits{J.}},
\bauthor{\bsnm{Ghosh}, \binits{S.}},
\bauthor{\bsnm{Shams-Ansari}, \binits{A.}},
\bauthor{\bsnm{Puma}, \binits{E.}},
\bauthor{\bsnm{Sinclair}, \binits{N.}},
\bauthor{\bsnm{Reimer}, \binits{C.}},
\bauthor{\bsnm{Zhang}, \binits{M.}},
\bauthor{\bsnm{Lon{\v{c}}ar}, \binits{M.}}:
\batitle{{Integrated photonics on thin-film lithium niobate}}.
\bjtitle{Advances in Optics and Photonics}
\bvolume{13}(\bissue{2}),
\bfpage{242}
(\byear{2021})
\doiurl{10.1364/AOP.411024}
\end{barticle}
\endbibitem

\bibitem[\protect\citeauthoryear{Hu et~al.}{2024}]{Hu2024}
\begin{botherref}
\oauthor{\bsnm{Hu}, \binits{Y.}},
\oauthor{\bsnm{Zhu}, \binits{D.}},
\oauthor{\bsnm{Lu}, \binits{S.}},
\oauthor{\bsnm{Zhu}, \binits{X.}},
\oauthor{\bsnm{Song}, \binits{Y.}},
\oauthor{\bsnm{Renaud}, \binits{D.}},
\oauthor{\bsnm{Assumpcao}, \binits{D.}},
\oauthor{\bsnm{Cheng}, \binits{R.}},
\oauthor{\bsnm{Xin}, \binits{C.}},
\oauthor{\bsnm{Yeh}, \binits{M.}},
\oauthor{\bsnm{Warner}, \binits{H.}},
\oauthor{\bsnm{Guo}, \binits{X.}},
\oauthor{\bsnm{Shams-Ansari}, \binits{A.}},
\oauthor{\bsnm{Barton}, \binits{D.}},
\oauthor{\bsnm{Sinclair}, \binits{N.}},
\oauthor{\bsnm{Loncar}, \binits{M.}}:
{Integrated electro-optics on thin-film lithium niobate}.
ArXiv,
9--12
(2024)
{\href{https://arxiv.org/abs/2404.06398}{{arXiv:2404.06398}}}
\end{botherref}
\endbibitem

\bibitem[\protect\citeauthoryear{Wang et~al.}{2024}]{Wang2024}
\begin{barticle}
\bauthor{\bsnm{Wang}, \binits{C.}},
\bauthor{\bsnm{Li}, \binits{Z.}},
\bauthor{\bsnm{Riemensberger}, \binits{J.}},
\bauthor{\bsnm{Lihachev}, \binits{G.}},
\bauthor{\bsnm{Churaev}, \binits{M.}},
\bauthor{\bsnm{Kao}, \binits{W.}},
\bauthor{\bsnm{Ji}, \binits{X.}},
\bauthor{\bsnm{Zhang}, \binits{J.}},
\bauthor{\bsnm{Blesin}, \binits{T.}},
\bauthor{\bsnm{Davydova}, \binits{A.}},
\bauthor{\bsnm{Chen}, \binits{Y.}},
\bauthor{\bsnm{Huang}, \binits{K.}},
\bauthor{\bsnm{Wang}, \binits{X.}},
\bauthor{\bsnm{Ou}, \binits{X.}},
\bauthor{\bsnm{Kippenberg}, \binits{T.J.}}:
\batitle{{Lithium tantalate photonic integrated circuits for volume manufacturing}}.
\bjtitle{Nature}
\bvolume{629}(\bissue{8013}),
\bfpage{784}--\blpage{790}
(\byear{2024})
\doiurl{10.1038/s41586-024-07369-1}
\end{barticle}
\endbibitem

\end{thebibliography}



\begin{thebibliography}{9}
\ifx \bisbn   \undefined \def \bisbn  #1{ISBN #1}\fi
\ifx \binits  \undefined \def \binits#1{#1}\fi
\ifx \bauthor  \undefined \def \bauthor#1{#1}\fi
\ifx \batitle  \undefined \def \batitle#1{#1}\fi
\ifx \bjtitle  \undefined \def \bjtitle#1{#1}\fi
\ifx \bvolume  \undefined \def \bvolume#1{\textbf{#1}}\fi
\ifx \byear  \undefined \def \byear#1{#1}\fi
\ifx \bissue  \undefined \def \bissue#1{#1}\fi
\ifx \bfpage  \undefined \def \bfpage#1{#1}\fi
\ifx \blpage  \undefined \def \blpage #1{#1}\fi
\ifx \burl  \undefined \def \burl#1{\textsf{#1}}\fi
\ifx \doiurl  \undefined \def \doiurl#1{\url{https://doi.org/#1}}\fi
\ifx \betal  \undefined \def \betal{\textit{et al.}}\fi
\ifx \binstitute  \undefined \def \binstitute#1{#1}\fi
\ifx \binstitutionaled  \undefined \def \binstitutionaled#1{#1}\fi
\ifx \bctitle  \undefined \def \bctitle#1{#1}\fi
\ifx \beditor  \undefined \def \beditor#1{#1}\fi
\ifx \bpublisher  \undefined \def \bpublisher#1{#1}\fi
\ifx \bbtitle  \undefined \def \bbtitle#1{#1}\fi
\ifx \bedition  \undefined \def \bedition#1{#1}\fi
\ifx \bseriesno  \undefined \def \bseriesno#1{#1}\fi
\ifx \blocation  \undefined \def \blocation#1{#1}\fi
\ifx \bsertitle  \undefined \def \bsertitle#1{#1}\fi
\ifx \bsnm \undefined \def \bsnm#1{#1}\fi
\ifx \bsuffix \undefined \def \bsuffix#1{#1}\fi
\ifx \bparticle \undefined \def \bparticle#1{#1}\fi
\ifx \barticle \undefined \def \barticle#1{#1}\fi
\bibcommenthead
\ifx \bconfdate \undefined \def \bconfdate #1{#1}\fi
\ifx \botherref \undefined \def \botherref #1{#1}\fi
\ifx \url \undefined \def \url#1{\textsf{#1}}\fi
\ifx \bchapter \undefined \def \bchapter#1{#1}\fi
\ifx \bbook \undefined \def \bbook#1{#1}\fi
\ifx \bcomment \undefined \def \bcomment#1{#1}\fi
\ifx \oauthor \undefined \def \oauthor#1{#1}\fi
\ifx \citeauthoryear \undefined \def \citeauthoryear#1{#1}\fi
\ifx \endbibitem  \undefined \def \endbibitem {}\fi
\ifx \bconflocation  \undefined \def \bconflocation#1{#1}\fi
\ifx \arxivurl  \undefined \def \arxivurl#1{\textsf{#1}}\fi
\csname PreBibitemsHook\endcsname

\bibitem[\protect\citeauthoryear{Lange et~al.}{2022}]{Lange2022}
\begin{barticle}
\bauthor{\bsnm{Lange}, \binits{N.A.}},
\bauthor{\bsnm{H{\"{o}}pker}, \binits{J.P.}},
\bauthor{\bsnm{Ricken}, \binits{R.}},
\bauthor{\bsnm{Quiring}, \binits{V.}},
\bauthor{\bsnm{Eigner}, \binits{C.}},
\bauthor{\bsnm{Silberhorn}, \binits{C.}},
\bauthor{\bsnm{Bartley}, \binits{T.J.}}:
\batitle{{Cryogenic integrated spontaneous parametric down-conversion}}.
\bjtitle{Optica}
\bvolume{9}(\bissue{1}),
\bfpage{108}
(\byear{2022})
\doiurl{10.1364/optica.445576}
{\href{https://arxiv.org/abs/2110.07425}{{arXiv:2110.07425}}}
\end{barticle}
\endbibitem

\bibitem[\protect\citeauthoryear{Lange et~al.}{2023}]{Lange2023}
\begin{barticle}
\bauthor{\bsnm{Lange}, \binits{N.A.}},
\bauthor{\bsnm{Schapeler}, \binits{T.}},
\bauthor{\bsnm{H{\"{o}}pker}, \binits{J.P.}},
\bauthor{\bsnm{Protte}, \binits{M.}},
\bauthor{\bsnm{Bartley}, \binits{T.J.}}:
\batitle{{Degenerate photons from a cryogenic spontaneous parametric down-conversion source}}.
\bjtitle{Physical Review A}
\bvolume{108}(\bissue{2}),
\bfpage{023701}
(\byear{2023})
\doiurl{10.1103/PhysRevA.108.023701}
{\href{https://arxiv.org/abs/2303.17428}{{arXiv:2303.17428}}}
\end{barticle}
\endbibitem

\bibitem[\protect\citeauthoryear{Thiele et~al.}{2020}]{Thiele2020}
\begin{barticle}
\bauthor{\bsnm{Thiele}, \binits{F.}},
\bauthor{\bsnm{Bruch}, \binits{F.}},
\bauthor{\bsnm{Quiring}, \binits{V.}},
\bauthor{\bsnm{Ricken}, \binits{R.}},
\bauthor{\bsnm{Herrmann}, \binits{H.}},
\bauthor{\bsnm{Eigner}, \binits{C.}},
\bauthor{\bsnm{Silberhorn}, \binits{C.}},
\bauthor{\bsnm{Bartley}, \binits{T.J.}}:
\batitle{{Cryogenic electro-optic polarisation conversion in titanium in-diffused lithium niobate waveguides}}.
\bjtitle{Optics Express}
\bvolume{28}(\bissue{20}),
\bfpage{28961}
(\byear{2020})
\doiurl{10.1364/OE.399818}
\end{barticle}
\endbibitem

\bibitem[\protect\citeauthoryear{Thiele et~al.}{2022}]{Thiele2022}
\begin{barticle}
\bauthor{\bsnm{Thiele}, \binits{F.}},
\bauthor{\bsnm{{Vom Bruch}}, \binits{F.}},
\bauthor{\bsnm{Brockmeier}, \binits{J.}},
\bauthor{\bsnm{Protte}, \binits{M.}},
\bauthor{\bsnm{Hummel}, \binits{T.}},
\bauthor{\bsnm{Ricken}, \binits{R.}},
\bauthor{\bsnm{Quiring}, \binits{V.}},
\bauthor{\bsnm{Lengeling}, \binits{S.}},
\bauthor{\bsnm{Herrmann}, \binits{H.}},
\bauthor{\bsnm{Eigner}, \binits{C.}},
\bauthor{\bsnm{Silberhorn}, \binits{C.}},
\bauthor{\bsnm{Bartley}, \binits{T.J.}}:
\batitle{{Cryogenic electro-optic modulation in titanium in-diffused lithium niobate waveguides}}.
\bjtitle{JPhys Photonics}
\bvolume{4}(\bissue{3}),
\bfpage{28961}--\blpage{28968}
(\byear{2022})
\doiurl{10.1088/2515-7647/ac6c63}
{\href{https://arxiv.org/abs/2202.00306}{{2202.00306}}}
\end{barticle}
\endbibitem

\bibitem[\protect\citeauthoryear{Thiele et~al.}{2023}]{Thiele2023}
\begin{barticle}
\bauthor{\bsnm{Thiele}, \binits{F.}},
\bauthor{\bsnm{Hummel}, \binits{T.}},
\bauthor{\bsnm{McCaughan}, \binits{A.N.}},
\bauthor{\bsnm{Brockmeier}, \binits{J.}},
\bauthor{\bsnm{Protte}, \binits{M.}},
\bauthor{\bsnm{Quiring}, \binits{V.}},
\bauthor{\bsnm{Lengeling}, \binits{S.}},
\bauthor{\bsnm{Eigner}, \binits{C.}},
\bauthor{\bsnm{Silberhorn}, \binits{C.}},
\bauthor{\bsnm{Bartley}, \binits{T.J.}}:
\batitle{{All optical operation of a superconducting photonic interface}}.
\bjtitle{Optics Express}
\bvolume{31}(\bissue{20}),
\bfpage{32717}
(\byear{2023})
\doiurl{10.1364/oe.492035}
{\href{https://arxiv.org/abs/2302.12123}{{2302.12123}}}
\end{barticle}
\endbibitem

\bibitem[\protect\citeauthoryear{Wong et~al.}{2020}]{Wong2020}
\begin{barticle}
\bauthor{\bsnm{Wong}, \binits{W.T.}},
\bauthor{\bsnm{Hosseini}, \binits{M.}},
\bauthor{\bsnm{Rucker}, \binits{H.}},
\bauthor{\bsnm{Bardin}, \binits{J.C.}}:
\batitle{{A 1 mW cryogenic LNA exploiting optimized SiGe HBTs to achieve an average noise temperature of 3.2 K from 4-8 GHz}}.
\bjtitle{IEEE MTT-S International Microwave Symposium Digest}
\bvolume{2020-Augus},
\bfpage{181}--\blpage{184}
(\byear{2020})
\doiurl{10.1109/IMS30576.2020.9224049}
\end{barticle}
\endbibitem

\bibitem[\protect\citeauthoryear{Avenhaus et~al.}{2008}]{Avenhaus2008}
\begin{barticle}
\bauthor{\bsnm{Avenhaus}, \binits{M.}},
\bauthor{\bsnm{Coldenstrodt-Ronge}, \binits{H.B.}},
\bauthor{\bsnm{Laiho}, \binits{K.}},
\bauthor{\bsnm{Mauerer}, \binits{W.}},
\bauthor{\bsnm{Walmsley}, \binits{I.A.}},
\bauthor{\bsnm{Silberhorn}, \binits{C.}}:
\batitle{{Photon number statistics of multimode parametric down-conversion}}.
\bjtitle{Physical Review Letters}
\bvolume{101}(\bissue{5}),
\bfpage{1}--\blpage{4}
(\byear{2008})
\doiurl{10.1103/PhysRevLett.101.053601}
{\href{https://arxiv.org/abs/0804.0740}{{0804.0740}}}
\end{barticle}
\endbibitem

\bibitem[\protect\citeauthoryear{{Migdall, Alan and Polyakov, Sergey V and Fan, Jingyun and Bienfang}}{2013}]{Migdall2013}
\begin{bbook}
\bauthor{\bsnm{{Migdall, Alan and Polyakov, Sergey V and Fan, Jingyun and Bienfang}}, \binits{J.C.}}:
\bbtitle{Single-Photon Generation and Detection: Physics and Applications},
\bedition{45} edn.
\bpublisher{Academic Press}, \blocation{???}
(\byear{2013}).
\doiurl{10.1016/B978-0-12-387695-9.00002-0} .
\burl{https://www.sciencedirect.com/bookseries/experimental-methods-in-the-physical-sciences/vol/45/suppl/C https://linkinghub.elsevier.com/retrieve/pii/B9780123876959000020}
\end{bbook}
\endbibitem

\bibitem[\protect\citeauthoryear{Miatto et~al.}{2018}]{Miatto2018}
\begin{barticle}
\bauthor{\bsnm{Miatto}, \binits{F.M.}},
\bauthor{\bsnm{Safari}, \binits{A.}},
\bauthor{\bsnm{Boyd}, \binits{R.W.}}:
\batitle{{Explicit formulas for photon number discrimination with on/off detectors}}.
\bjtitle{Applied Optics}
\bvolume{57}(\bissue{23}),
\bfpage{6750}
(\byear{2018})
\doiurl{10.1364/ao.57.006750}
\end{barticle}
\endbibitem

\end{thebibliography}

\end{document}


\title[Supplementary Material: Cryogenic feed-forward of a Photonic Quantum State]{Supplementary Material: Cryogenic feed-forward of a Photonic Quantum State}


\author*[1,2]{\fnm{Frederik} \sur{Thiele}}\email{frederik.thiele@uni-paderborn.de}
\equalcont{These authors contributed equally to this work.}

\author[2]{\fnm{Niklas} \sur{Lamberty}}
\equalcont{These authors contributed equally to this work.}

\author[1]{\fnm{Thomas} \sur{Hummel}}

\author[1,2]{\fnm{Nina A.} \sur{Lange}}

\author[1,2]{\fnm{Lorenzo M.} \sur{Procopio}}

\author[1,2]{\fnm{Aishi} \sur{Barua}}

\author[3]{\fnm{Sebastian} \sur{Lengeling}}

\author[1]{\fnm{Viktor} \sur{Quiring}}

\author[1]{\fnm{Christof} \sur{Eigner}}

\author[3]{\fnm{Christine} \sur{Silberhorn}}

\author[1,2]{\fnm{Tim J.} \sur{Bartley}}

\affil[1]{\orgdiv{Institute for Photonic Quantum Systems (PhoQS)}, \orgname{Paderborn University}, \orgaddress{\street{Warburger Str. 100}, \city{Paderborn}, \postcode{33098},  \country{Germany}}}

\affil[2]{\orgdiv{Department of Physics}, \orgname{Paderborn University}, \orgaddress{\street{Warburger Str. 100}, \city{Paderborn}, \postcode{33098},  \country{Germany}}}

\affil[3]{\orgdiv{Integrated Quantum Optics Group, Institute for Photonic Quantum Systems (PhoQS)}, \orgname{Paderborn University}, \orgaddress{\street{Warburger Str. 100}, \city{Paderborn}, \postcode{33098},  \country{Germany}}}

\keywords{feedforward, quantum photonics, single-photon, cryogenic}

\maketitle
\section{Photonic setup for feedforward}
\begin{figure}
        \centering
        \includegraphics[width=0.8\linewidth]{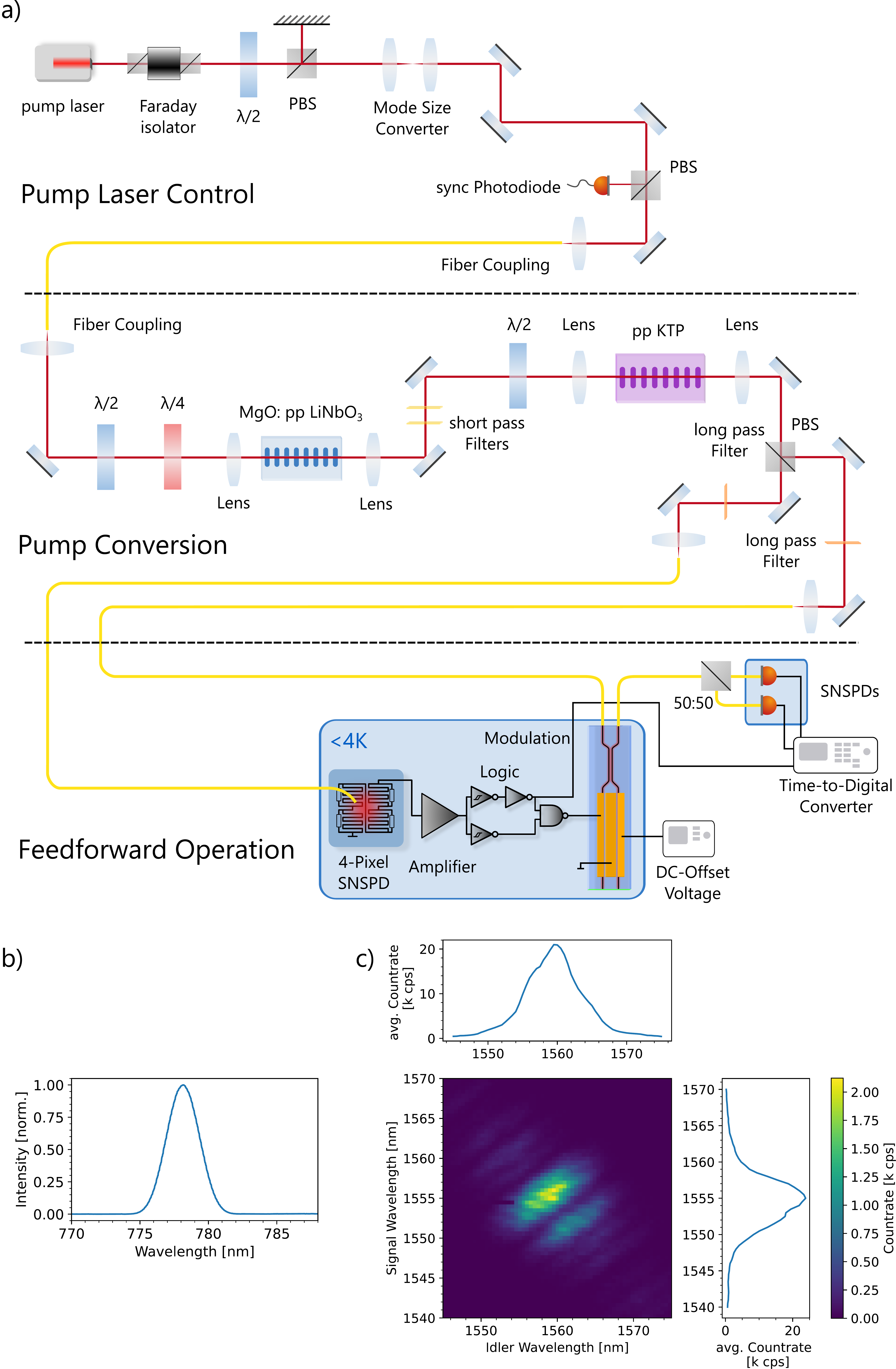}
        \caption{a) Layout of the photonic circuit to realize the feedforward operation of a quantum state. The pump power is controlled in first section. In the second section the pump light is up-converted with SHG to \SI{778.1}{nm} and used to generate the correlated SPDC photons at \SIrange{1555}{1560}{nm}. In the bottom section, the feedforward operation is shown. b) Resulting spectrum after the second harmonic generation with the initial pump laser. c) Joint spectral intensity distribution after the parametric down-conversion. The inset graphs show the column and row average for the idler and signal photons.}
        \label{fig:EntireLayout}
    \end{figure}

The complete feedforward operation using quantum states involves state generation and the subsequent manipulation. In the following, we outline the layout of the photonic and electronic systems. Our implementation comprises three primary sections: pump laser control, pump conversion for generating photon pairs with Spontaneous Parametric Down-Conversion (SPDC), and the main feedforward operation, as shown in Fig.~\ref{fig:EntireLayout}~a). 
    
The quantum states for the feedforward operation are generated in a SPDC-source. To generate photons pairs in the technologically relevant telecom C-band, we need input light around \SI{775}{nm}. To do so, we use a pulsed laser with a central frequency around \SI{1560}{nm}. The light is then up-converted via second harmonic generation (SHG) to around \SI{780}{nm}. In the subsequent SPDC process, photon pairs are generated with a wavelength in the range of \SIrange{1555}{1560}{nm}. 

The pump laser has a peak wavelength at \SI{1557.4}{nm}, a bandwidth of \SI{13.8}{nm} (FWHM) and generates \SI{191}{fs} pulses with a \SI{12.5}{ns} repetition period (NKT photonics Origami). To control the power of the pump laser, we couple the linearly polarized light through a rotatable half wave plate and polarizing beamsplitter (PBS). The light is then coupled into a single mode fiber and guided towards the SHG crystal. To correct polarization rotations in the single mode fiber, we couple the light out of the fiber and transmit it through a half- and quarter waveplate. The pump light is then up-converted with a periodically poled lithium niobate crystal doped with magnesium oxide (Covesion). To stabilize the output spectrum, the SHG-crystal is mounted in a temperature-controlled oven. The resulting up-converted spectrum has central wavelength of \SI{778.1}{nm} with a bandwidth of \SI{2.8}{nm}, as shown in Fig.~\ref{fig:EntireLayout}~b). Afterwards, the pump light is filtered out with a short pass filter and the polarization rotated with a half waveplate. 

The up-converted light is then transmitted into a periodically poled Potassium Titanyl Phosphate (ppKTP) crystal (AdvR), where it generates correlated photon pairs within the ppKTP waveguide. As result correlated photon pairs are generated with a joint spectral distribution around \SI{1555}{nm}, as shown in Fig.\ref{fig:EntireLayout}~c). The idler spectrum peaks at \SI{1559.5}{nm} and the signal spectrum at \SI{1555}{nm}, both having a full width at half maximum (FWHM) of \SI{8}{nm}. The joint spectral intensity was obtained using a pair of home-built single-photon sensitive grating spectrometers~\cite{Lange2022,Lange2023}.

A PBS is used to split the correlated photons produced by the Type-II SPDC process based on their polarization, while any remaining pump light is removed using a long-pass filter. The SPDC photons are coupled into a single-mode fiber for the feedforward process.  
 
For the feedforward process, idler photons are directed to a commercial four-pixel Superconducting Nanowire Single-Photon Detector (SNSPD) (Photon Spot), which generates a detection signal. This signal is processed by an amplifier and a logic circuit, which discriminates the signal amplitude to determine the number of incident photons. More details on the amplifier and discriminator circuit are provided in section~\ref{sec:elec}. The output signal from the logic circuit is sent to the electrodes of the modulator.    
   
Signal photons are guided via single-mode fibers to an electro-optic modulator. Depending on the photon number detected in the idler arm, the signal photons are either transmitted through the modulator or suppressed. Further details on the electro-optic modulator are discussed in section~\ref{sec:Modulator}. The transmitted signal photons are analyzed using a Hanbury-Brown-Twiss (HBT) detection setup. A 50:50 beam splitter divides the signal photons into two paths and are detected by SNSPDs.    

A time-to-digital converter captures the detection signals for further analysis. During acquisition, both the trigger signals from the logic circuit and the HBT-SNSPD signals are collected for processing. The HBT-SNSPDs (Photon Spot) and time-to-digital converter are commercially available devices (Swabian Instruments).

\section{Modulator}\label{sec:Modulator}

\begin{figure}
    \centering
    \includegraphics[width=0.75\linewidth]{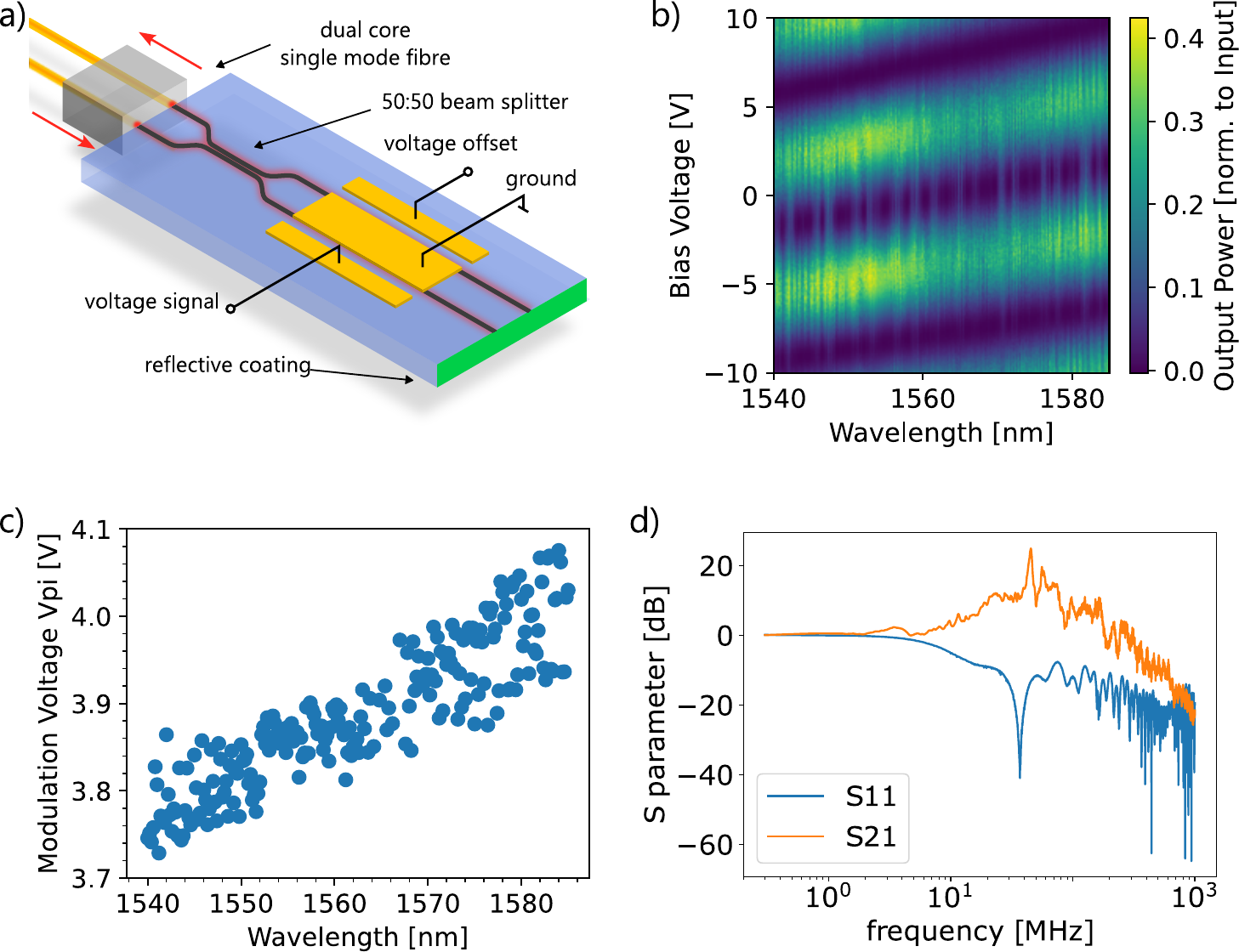}
    \caption{a) Layout of the fiber coupled electro-optic modulator in the Michelson-interferometer configuration. Titanium in-diffused waveguides transmit the light in a z-cut lithium niobate chip. The single mode fibers are attached with a UV-curable glue. b) Bias-voltage and wavelength sweep of the electro-optic modulator, at \SI{4}{K}. c) Extracted modulation voltage $V_{\pi}$ of the modulator to switch between the minimum and maximum in the output intensity, at \SI{4}{K}. d) The S-parameters of the modulator measured at 4~K. The S-parameters describe the electrical reflection (S11) at one electrode pair which is contacted by wire bonds. The electric-to-optical conversion is described by the S21-parameter with an input signal at the electrode and an optical readout of transmitted light acquired with a photodiode. The dataset is not corrected for photodiode sensitivity and cable transmission inside the cryostat. The S21 is normalized for low frequencies.}
    \label{fig:ModChar}
\end{figure}

The cryogenic feedforward operation requires a manipulation of photons. In this implementation, the signal photons are modulated with phase modulators integrated in a Michelson interferometer. To do so, light is guided from single mode fibers into waveguides in titanium in-diffused lithium niobate. The Michelson interferometer is formed by an integrated  50:50 beam splitter, phase modulators and reflector at the end-facet, as shown in Fig.~\ref{fig:ModChar}~a). Incoming light is split into two paths, the light experiences a voltage dependent phase-shift from the phase modulator and is reflected by a high reflectivity coating at the end-facet, before undergoing the same phase shift in the reverse direction. The two paths then recombine at the same 50:50 integrated beam splitter; dependent on the total acquired phase, the light will be exit each port with different amplitude. The Michelson interferometer and integrated phase-shifter thus acts as a reconfigurable mirror. 

The integrated waveguide platform is realized by titanium in-diffusion in z-cut lithium niobate~\cite{Thiele2020,Thiele2022}. This modulator layout has \SI{25}{mm} long electrodes, with a electrode separation of \SI{11}{\micro m}. A Signal-Ground-Signal electrode geometry is used: the middle electrode is grounded, one electrode is used for the modulation signal and the third electrode is used for the bias off-set. The beam splitter has a total length of \SI{15}{mm} including the bends. The reflective end-facet coating achieves a reflectivity of 97\%. The overall modulator layout has been previously tested by Thiele et. al.~\cite{Thiele2023}. 

Light is transmitted in and out of the waveguides by a single fiber-pigtail containing two single mode fibers (SMF28). To do so, the fiber-pigtail is attached with a UV-curable adhesive (Norland 81) at the end-facet of the chip. The theoretical mode overlap between a single mode fiber and the waveguide is 90\%. At room temperature, the total throughput of the device was 27\%. At cryogenic temperatures, we achieved a maximal fiber-to-fiber transmission through entire modulator of 8\%. The reduction of the transmission mainly arises by thermal stress-induced coupling changes at the fiber interface during cooling.

The modulator is characterized by sweeping the operation wavelength with narrow-band laser (\SI{1}{pm}) and sweeping the input voltage at one electrode, as shown in Fig.~\ref{fig:ModChar}~b). As a result, a modulation voltage can be extracted, as shown in Fig.~\ref{fig:ModChar}~c). The modulation voltage is defined as the voltage required to change the output intensity from a minimum to a maximum, the ratio of which defines the extinction ratio. In this modulator, an extinction ratio of \SI{18}{dB} was achieved using a narrowband CW laser (EXFO). In the feedforward experiment, the modulator is operated with pulsed SPDC light which reduces the extinction ratio to \SI{10.2(0.2)}{dB} due to the broader bandwidth.

The modulation signal is transmitted to the electrodes through cryogenic coax lines and wire bonds. In this layout the electrode structure can be seen as a capacitor. To test the optical-to-electrical modulation bandwidth we connected a vector network analyzer to the cryogenic modulator. The electrical port (1) are the electrodes of the modulator. The optical performance (2) is acquired by illuminating the modulator at \SI{1556}{nm} and acquiring the optical modulation with a photodiode. As a result, we acquired a \SI{3}{dB}-bandwidth of \SI{230}{MHz}, as shown in Fig.\ref{fig:ModChar}~d).  

\section{Electronics}\label{sec:elec}

\begin{figure}
    \centering
    \includegraphics[width=1\linewidth]{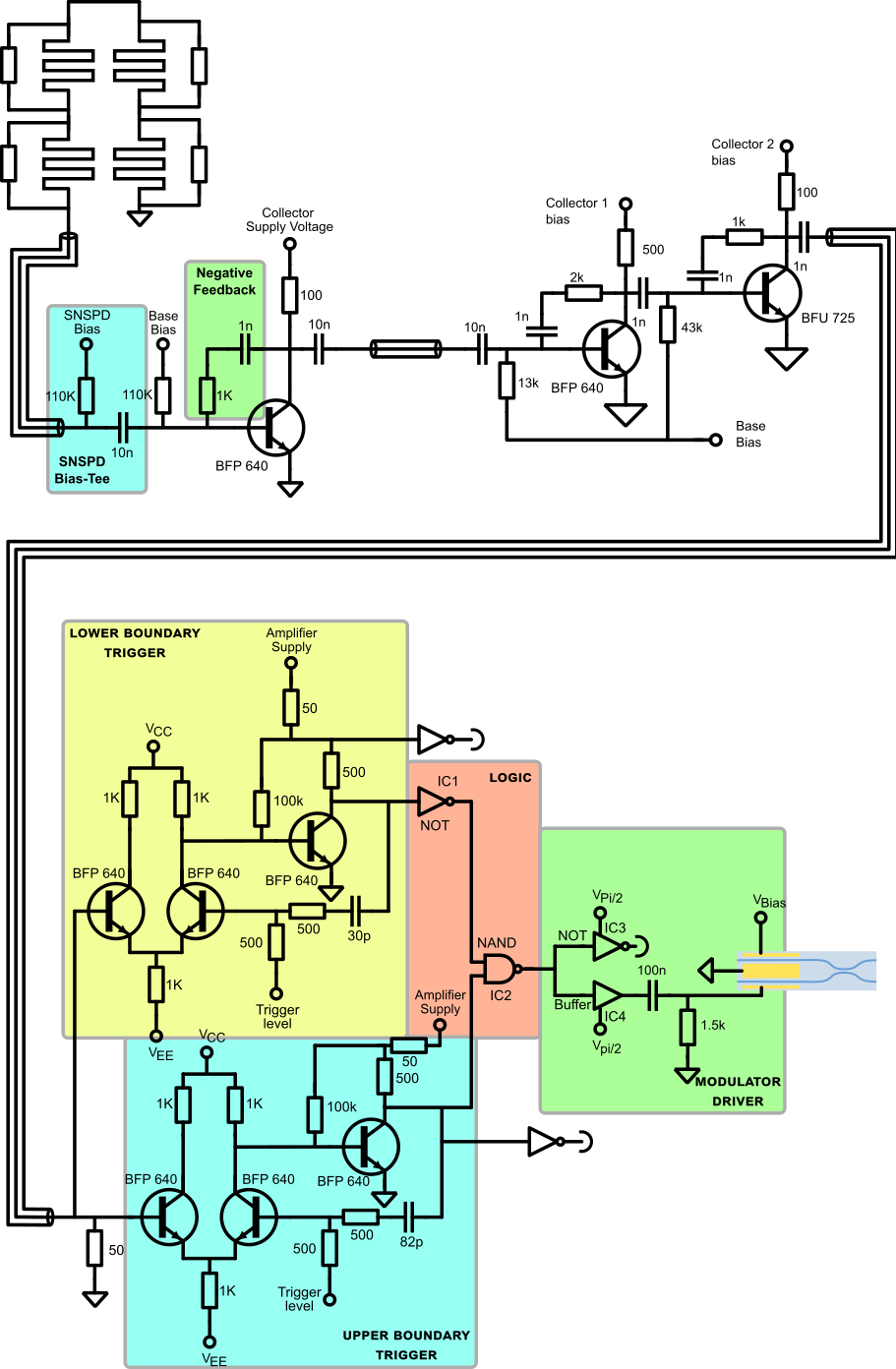}
    \caption{The complete electronic circuit used to perform the feed-forward operation. DC-bypass capacitors are omitted, but are present on all constant voltages, with 1nF and/or 100nF.}
    \label{fig:circuit}
\end{figure}

The electronics used to perform the feed-forward operation are based on custom made printed circuit boards (PCBs) with discrete components. The components used are thin film NiCr resistors, NP0 capacitors, SiGe heterojunction bipolar transistors (HBTs) and CMOS logic gates. These active components are in principle compatible with integration in a SiGe BiCMOS process, where fabrication processes optimized for cryogenic temperature are available~\cite{Wong2020}. A list of components used in this circuit can be found in table \ref{tab:BOM} and the full circuit diagram can be seen in figure \ref{fig:circuit}.

\begin{figure}
        \centering
        \includegraphics[width=0.7\linewidth]{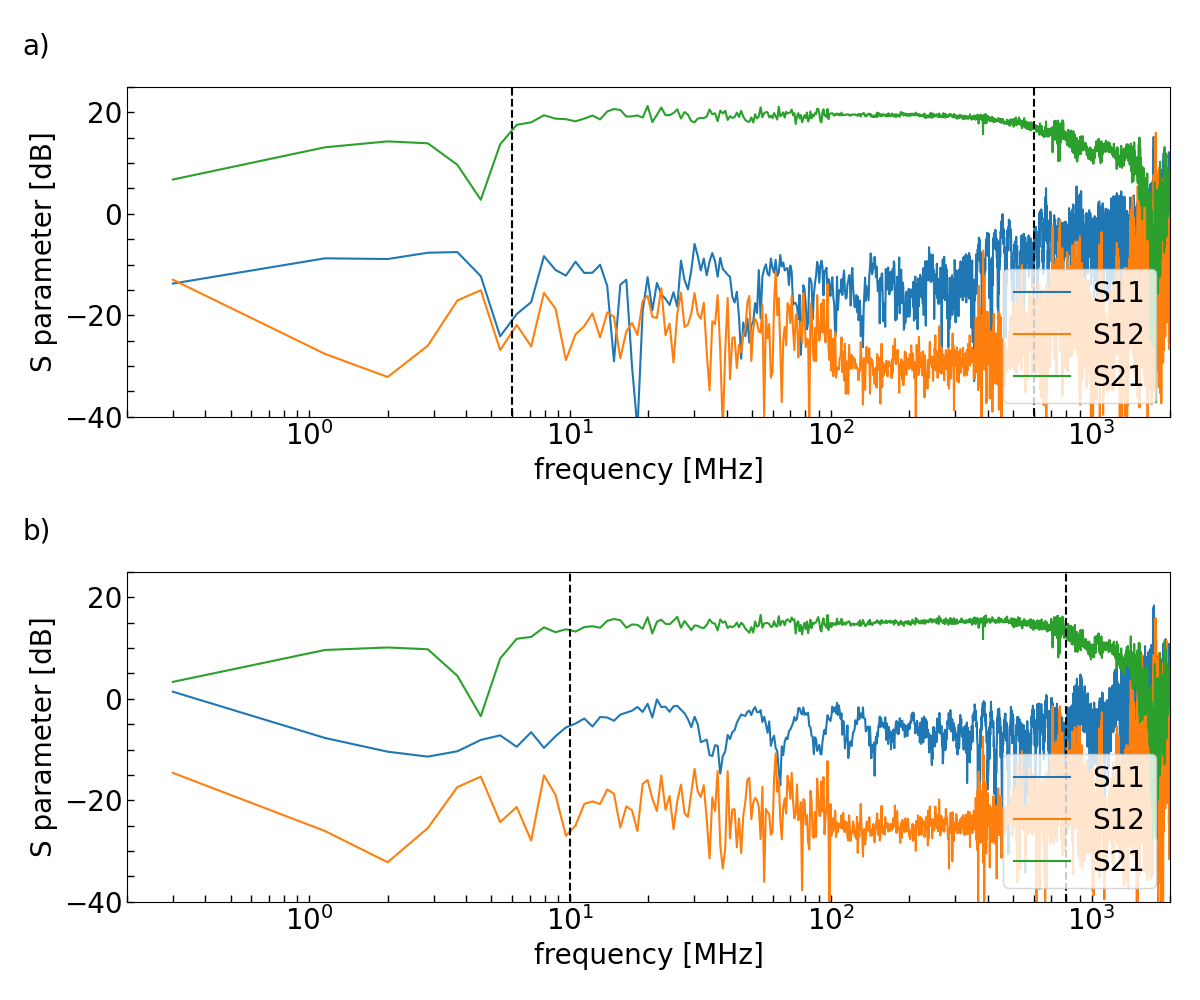}
        \caption{The S-parameters of the first stage amplifier for a) $\sim$50 $\mu$A Base current and 0.6 V at the collector and b) $\sim$17 $\mu$A Base current and 0.4 V at the collector. The operating point used in this experiment lies between these two points. The S-parameters can be interpreted as: S11-Reflection at the input; S21-Amplification from input to output and S12-leakage from the output back to the input.}
        \label{fig:Amp1}
    \end{figure}

\begin{figure}
        \centering
        \includegraphics[width=0.8\linewidth]{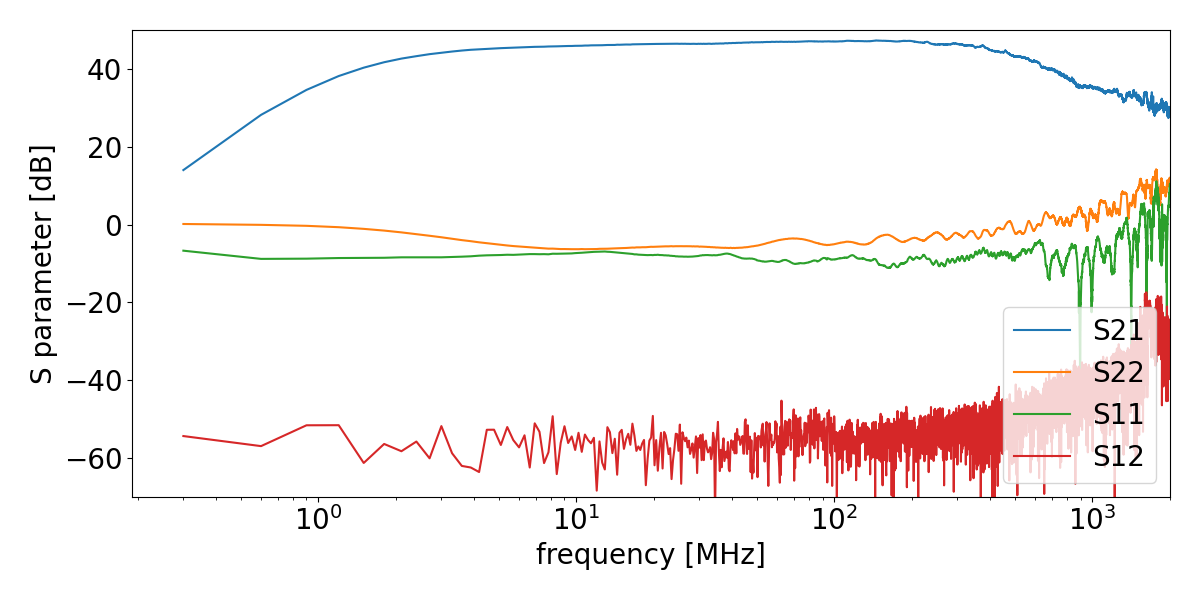}
        \caption{The S-parameters of the second stage amplifier. Tested at 4~K with an input power of -40~dBm. The S-parameters can be interpreted as:  S21-Amplification from input to output; S22-Reflection at the output; S11-Reflection at the input; and S12-leakage from the output back to the input.}
        \label{fig:Amp2}
    \end{figure}

The full electronic feed-forward is comprised of three different PCBs, two amplifiers and one waveform analyser. The amplifiers are based on a low noise amplifier architecture. They feature an additional resistor-capacitor feedback loop to stabilise the amplifier and prevent oscillations. The amplifiers are operated in weak saturation to reduce the power consumption. These amplifiers are needed to boost the signal from the sub mV range to about 100 mV for further height analysis. Analysis of the scattering parameters (S-parameters) has been carried out at \SI{4}{K} to characterise the amplifiers. The results of this can be seen in figures \ref{fig:Amp1} and \ref{fig:Amp2}.
The first amplifier achieves a bandwidth of \SI{6}{MHz} to \SI{600}{MHz}, which covers most of the SNSPDs signal spectrum and thus minimizes signal distortions. The input reflectivity of the amplifier is consistently below \SI{-10}{dB} in the relevant frequency band for the SNSPD signal. This protects the SNSPD from being disturbed by signal reflections. Having a low leakage from the output of the amplifier back to the input is also beneficial, as noise generated by subsequent circuits will be attenuated before reaching the SNSPD.

For the feed-forward circuit shown in this work we require more than simple signal amplification. The aim of the designed logic circuit is to generate modulator switching signals based on measuring a given photon number. In order to achieve this photon number selection for the feed-forward circuit, the amplitude multiplexed SNSPD signal needs to be converted into digital signals indicating different detected photon numbers. A typical way to perform this transition from analog to digital signals is with a Schmitt-trigger. This circuit detects the crossing of a voltage threshold and switches a digital signal at the output based on this. 

Due to the limited availability of \SI{4}{K} compatible components, we designed a custom circuit to perform the role of a Schmitt-trigger. This requires a circuit that can detect the voltage difference between the signal and the triggering threshold, as well as a feedback loop to create a large output signal once this difference becomes positive.

As the differentiation circuit we use a differential amplifier, which is made up of two biploar transistors, which share a common current supply at the emitter. A change in voltage at the base of either transistor will increase the current in this transistor and thus decrease the current in the other transistor due to the shared current supply. The amplifiers therefore outputs the difference between the voltages at the transistor bases. This can be used to trigger on certain voltage levels by biassing one input at a variable voltage (Trigger Level). The amplifier will then only create a positive output voltage once this threshold voltage is crossed by the signal coming from the SNSPD.

The signal generated by this type of differentiating circuit is however not able to drive larger currents. We thus employ an additional low noise amplifier circuit to buffer the output and thus increase both the voltage and current generated by the trigger circuit. The stabilisation of the output amplifiers is achieved  with a common resistor for both base and collector bias. This acts as a negative feedback loop, required to stabilize the amplifier.

In order for this circuit to function as a Schmitt trigger, a way to introduce positive feedback between the output and the input of of the differential amplifier is required. The positive feedback causes a small positive output voltage on the differential amplifier to increase the voltage at the input and thus in turn increase the voltage at the output. This feedback loop continues until the output of the differential amplifier is saturated. It thus creates a large voltage signal from a small one. In this circuit the positive feedback is introduced with a resistor voltage divider, to achieve a 1:1 mixing of the amplified SNSPD signal and the positive feedback at the input. An additional capacitor is introduced into this feedback line. This capacitor charges after a trigger event and thus gradually reduces the amount of positive feedback voltage. It thereby limits the time for which the positive feedback can be active. This decouples the pulse length of the SNSPD from the digital output pulse and thus allows free choice over the length of the subsequent digital signal.

\begin{table}[]
\begin{tabular}{|l|l|l|l|}
\hline
\textbf{Components}& \textbf{Type}& \textbf{Part Number}& \textbf{Manufacturer}\\ \hline
Resistors  & all   & RN73-series       & TE-Connectivity    \\ \hline
Capacitors & 1nF   & GRT1555C1H102FA2D & Murata Electronics \\ \hline
Capacitors & 10nF  & GRM1555C1E103GE1D & Murata Electronics \\ \hline
Capacitors & 100nF & GCM31C5C1H104FA16 & Murata Electronics \\ \hline
Capacitors & 82p   & GRT0335C1H820FA2D & Murata Electronics \\ \hline
Capacitors & 30p   & GCM1555C1H300GA6D & Murata Electronics \\ \hline
Gate       & IC1   & NC7SP04P5X        & Onsemi             \\ \hline
Gate       & IC2   & NC7SP00P5X        & Onsemi             \\ \hline
Gate       & IC3   & NC7SV04P5X        & Onsemi             \\ \hline
Gate       & IC4   & NC7SV34P5X        & Onsemi             \\ \hline
\end{tabular}\label{tab:BOM}
\caption{A list of the components used to assemble the electronic circuits.}
\end{table}

    \begin{table}[]
\begin{tabular}{|l|l|l|l|} \hline 

\multicolumn{1}{|l|}{\textbf{Connection}}               & \multicolumn{1}{|l|}{\textbf{Current} {[}mA{]} ($\pm$0.01)} & \multicolumn{1}{|l|}{\textbf{Voltage} {[}V{]} ($\pm$0.01)} & \multicolumn{1}{|l|}{\textbf{Heatload} [mW]} \\ \hline 
\multicolumn{1}{|l|}{\textbf{Amplifier 1}}      &                                       \multicolumn{3}{|c|}{}\\ \hline  
\multicolumn{1}{|l|}{Base}                     & \multicolumn{1}{|r|}{0.02}             & \multicolumn{1}{|r|}{3.3}             & \multicolumn{1}{|l|}{}         \\ \hline  
\multicolumn{1}{|l|}{Collector}                & \multicolumn{1}{|r|}{1.65}             & \multicolumn{1}{|r|}{0.6}             & \multicolumn{1}{|r|}{1.1}      \\ \hline  
\multicolumn{1}{|l|}{\textbf{Amplifier 2}}     &                                       \multicolumn{3}{|c|}{}\\ \hline  
\multicolumn{1}{|l|}{Base}                     & \multicolumn{1}{|r|}{0.06}             & \multicolumn{1}{|r|}{1.68}            & \multicolumn{1}{|l|}{}         \\ \hline  
\multicolumn{1}{|l|}{Collector 1}              & \multicolumn{1}{|r|}{2.07}             & \multicolumn{1}{|r|}{1.4}             & \multicolumn{1}{|l|}{}         \\ \hline  
\multicolumn{1}{|l|}{Collector 2}              & \multicolumn{1}{|r|}{5.58}             & \multicolumn{1}{|r|}{1.4}             & \multicolumn{1}{|r|}{10.8}     \\ \hline  
\multicolumn{1}{|l|}{\textbf{Trigger}}         &                                       \multicolumn{3}{|c|}{}\\ \hline  
\multicolumn{1}{|l|}{Emitter}                  & \multicolumn{1}{|r|}{0.95}             & \multicolumn{1}{|r|}{-1.45}           & \multicolumn{1}{|l|}{}         \\ \hline  
\multicolumn{1}{|l|}{Collector}                & \multicolumn{1}{|r|}{0.88}             & \multicolumn{1}{|r|}{1.3}             & \multicolumn{1}{|r|}{2.5}      \\ \hline  
\multicolumn{1}{|l|}{\textbf{CMOS}}            &                                       \multicolumn{3}{|c|}{}\\ \hline  
\multicolumn{1}{|l|}{Logic / amplifier supply} & \multicolumn{1}{|r|}{1.7}              & \multicolumn{1}{|r|}{2.7}             & \multicolumn{1}{|l|}{}         \\ \hline  
\multicolumn{1}{|l|}{modulator driver}         & \multicolumn{1}{|r|}{4.4}              & \multicolumn{1}{|r|}{3.6}             & \multicolumn{1}{|r|}{20.4}     \\ \hline  
 & \multicolumn{2}{|c|}{}&\multicolumn{1}{|r|}{\textbf{Total:} \SI{34.8}{mW}}\\ \hline 
\end{tabular}
\caption{The operating voltages used in the experiment. Currents are shown for operating at 600 kHz detector count rate. The trigger voltages (not shown here) were adjusted between measurements to perform the desired photon number selection.}
\end{table}

\section{Simulation}

To validate the observed $g^{(2)}(0)$ values we simulate the effect of the feedforward circuit on the photon number distribution in the signal mode. For this we first compute the probability $p(n,n')$ for a given photon number $n$ to be detected as $n'$ photons by the SNSPD-array. This is computed with 
\begin{equation}
	p(n,n')=\Pi\cdot L_{j,i}(T),
\end{equation} 
where $\Pi$ is the Positive Operator Valued Measure (POVM) matrix of the detector, which accounts for the detector dependent association of a given incident photon number with a measured photon number, and $L_{j,i}$ is a loss matrix, which gives the probability for a given photon number to reach the detector. The loss matrix $L_{j,i}$ can be computed as
\begin{equation}
	L_{j,i}(T)=\binom{i}{j}\cdot T^i \cdot (1-T)^{(i-j)}~,
\end{equation}
where $i$ refers to the incident photon number, $j$ to the transmitted number of photons and $T$ to the coupling efficiency from the SPDC source to the detector.

Depending on which discrimination setting after the measurement is used, the columns $n'=1$ or $n'=2$ are chosen from the $p(n,n')$ matrix. The selected column gives the probability for $n$ photons in the idler mode to be detected as the chosen photon number and cause a modulation event. Since the idler and signal modes are correlated in the photon number, we can infer that idler's photon-number is equal to the signal's photon-number. To that end, we can simplify the simulation by only simulating only the photon-number distribution in the signal mode.

We want to compute the change induced by the feedforward circuit from a given input photon-number distribution $P(n)$ to a new distribution $P'(n)$. The initial photon-number distribution $P(n)$ is assumed to be Poissonian due to the multitude of Schmidt modes in our source (see figure \ref{fig:EntireLayout} c)). A high number of Schmidt modes causes the photon statistics to change from a thermal to Poissonian photon-number distribution \cite{Avenhaus2008}. To compute the resulting distribution $P'(n)$, we multiply the probability for a given photon number to generate a modulation signal $p(n,n'=1,2)$ with an inital distribution $P(n)$ such that;
\begin{equation}
	P'(n)= p(n,n'=1,2) \cdot P(n) ~.
\end{equation} 
In this simulation and experiments, we only consider events in which idler photons are detected. This selection reduces the vacuum component. In addition, we exclude events in which photons are detected at the HBT-detectors but no modulation has occurred, which is present due to a limited suppression ratio in the modulator.  The measured distribution consists only out of a photon number rightfully detected, cross-talk components of lower input photon number, and loss induced higher photon-number components.

We discard unheralded events by our photon-number discrimination. The resulting distribution $P'(n)$ is therefore renormalized to account for the discarding. 

From the resulting photon number distribution $P'(n)$ the $g^{(2)}(0)$ can be computed~\cite{Migdall2013} with
\begin{equation}
	g^{(2)} (0) = \frac{\sum_{n=0}^{\infty}n(n-1)P'(n)}{\left(\sum_{n=0}^{\infty}nP'(n)\right) ^2}~.
\end{equation}
The simulation allows us to compute the $g^{(2)}(0)$ for a sweep in the mean photon number created in the SPDC source.  The electronic discriminator allows us to select different ensembles of photon number for example; 2 photons or more than 2 photons. This can be simulated only selecting the columns in matrix $p(n,n')$ of the photon numbers selected by the discrimination circuit. The final detection probabilities are obtained by summing over each individual row.
    
\section{4-Pixel SNSPD}\label{sec:SNSPD}

\begin{figure}
    \centering
    \includegraphics[width=0.8\linewidth]{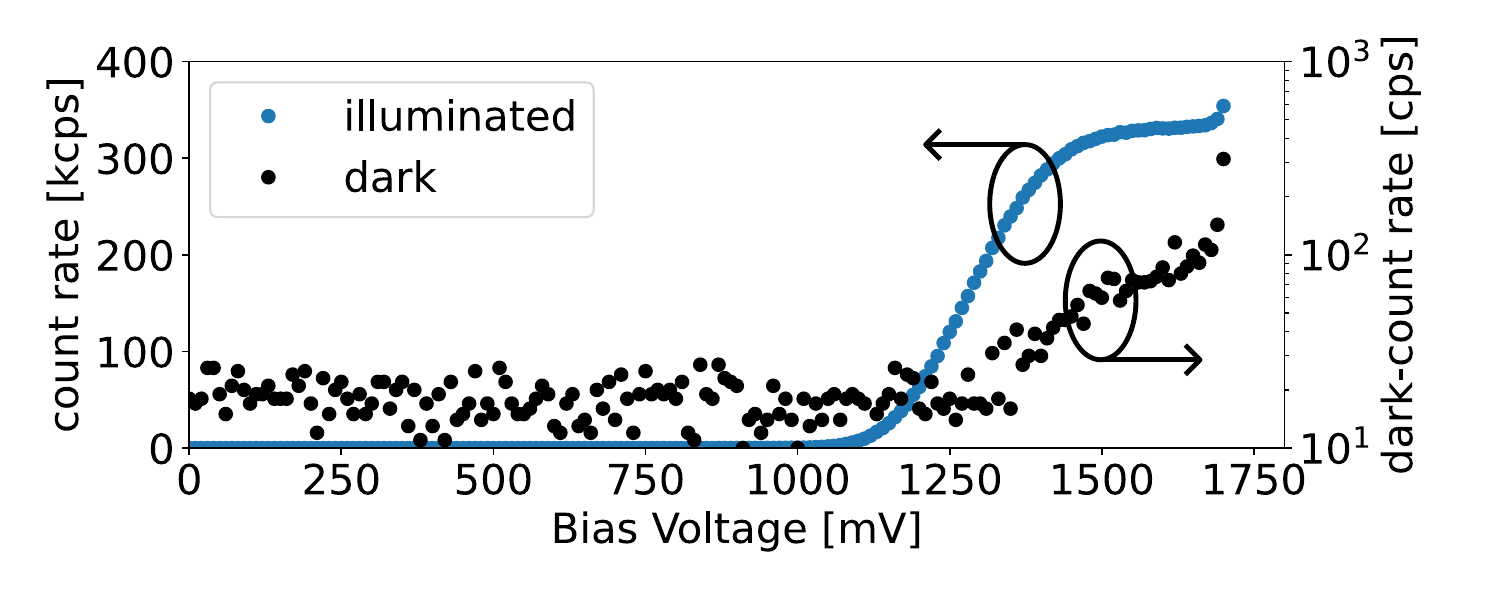}
    \caption{The count rate of the 4-pixel SNSPD under illumination and without illumination measured with all cryogenic electronics attached. The bias voltage biasses the SNSPD over a \SI{110}{k \ohm} resistor.}
    \label{fig:biassweep}
\end{figure}

In order to achieve the photon number resolved measurement required for our experiment we utilize a serial 4 pixel SNSPD array (Photonspot). This detector is made up of four SNSPD wired in series with a 50 $\Omega$ shunt resistor in parallel to each SNSPD. The detector is biased through a 110 k$\Omega$ resistor with a constant voltage supply. To show that the detector operation is unimpeded by the subsequent electronic processing, we perform a sweep of the bias current and measure the count rate recorded by our triggering circuit. The circuit is configured such that it will trigger on every photon number. The resulting change in count rate can be seen in figure \ref{fig:biassweep}. The detector reaches a plateau in the count rate, which indicates that the detector can be operated at a saturated internal detection efficiency, as is the case in normal operation.  After the plateau the detector experiences electro-thermal oscillations, which lead to a very high dark count rate. Due to the integrated \SI{50}{\ohm} resistors the detector will however not latch. The plateau voltage of \SI{1.6}{V} can be converted to an average bias current of \SI{14.5}{\micro A}, which is lower than the manufacturer provided optimal bias current of \SI{17}{\micro A}.

The dark-count rate of the detector rises for higher bias voltages as would be expected. However the dark-count rate does not fall to 0 for zero bias current, but remains at around \SI{20}{Hz}. This is likely caused by the trigger circuit triggering on noise in the cryostat, instead of detector signals. This additional dark-count rate is however small compared to dark counts in the plateau region of the detector, where it contributes only 10-20 \% of the total dark counts of around \SI{100}{Hz}.

\begin{table}[h]
\begin{tabular}{|l||l|l|l|l|l|}
\hline
   & 0     & 1     & 2     & 3     & 4     \\ \hline \hline
0  & 1.000 & 0.000 & 0.000 & 0.000 & 0.000 \\ \hline
1  & 0.300 & 0.682 & 0.017 & 0.000 & 0.000 \\ \hline
2  & 0.090 & 0.529 & 0.372 & 0.009 & 0.000 \\ \hline
3  & 0.027 & 0.313 & 0.519 & 0.139 & 0.003 \\ \hline
4  & 0.008 & 0.167 & 0.500 & 0.295 & 0.030 \\ \hline
5  & 0.002 & 0.085 & 0.412 & 0.417 & 0.084 \\ \hline
6  & 0.001 & 0.042 & 0.312 & 0.487 & 0.158 \\ \hline
7  & 0.000 & 0.020 & 0.225 & 0.510 & 0.245 \\ \hline
8  & 0.000 & 0.010 & 0.157 & 0.498 & 0.335 \\ \hline
9  & 0.000 & 0.005 & 0.107 & 0.465 & 0.423 \\ \hline
10 & 0.000 & 0.002 & 0.072 & 0.421 & 0.505 \\ \hline
\end{tabular}
\caption{The simulated POVM matrix of the 4 pixel SNSPD. The rows indicate the measured photon number while the columns indicate the incident photon number. The matrix shown here is cropped to a incident photon number of 10, but can be extended.}
\label{tab:POVM}
\end{table}

To simulate the response of the 4 pixel SNSPD, we compute a POVM matrix for the detector based on the observed crosstalk and previously measured detection efficiency. To compute the POVM we use the formulas given by Miatto et al. \cite{Miatto2018}. We additionally implement cross talk into the computed POVM matrix by reducing the probability for N=1,2,3 photon detection events proportionally by the crosstalk probability and add the deducted probability to the next higher photon number outcome. The resulting estimated POVM matrix can be seen in table \ref{tab:POVM}.

\section{Data availability}
All datasets used in this publication in the main manuscript and the supplementary material are openly available (\url{https://doi.org/10.5281/zenodo.13753477}).

\bibliography{sn-bibliography}